\documentclass[aps,prd,twocolumn,showpacs,showkeys]{revtex4-1}
\usepackage{graphicx}
\def\sig{{\cal I}}
\newcommand\g[1]{\mathrm{#1}}

\begin{document}
\title{Optimal observables for $Z'$ models in annihilation leptonic processes}
\author{A.V.~Gulov}
\email{alexey.gulov@gmail.com}
\affiliation{Dnipro National University, 49010 Dnipro, Ukraine}
\author{Ya.S.~Moroz}
\affiliation{Dnipro National University, 49010 Dnipro, Ukraine}
\begin{abstract}
The optimal observables with the best ratio of signal to statistical uncertainty are proposed for a bunch of popular models of the $Z'$ boson. They are the cross sections integrated over the phase space of the final particles with proper weight functions. It is shown that the proposed observables are completely equivalent to the $\chi^2$ fit of the differential cross section, so they could be used as an alternative of aggregating events into bins with further minimization of the $\chi^2$ function, especially in preliminary analysis of experimental data. Application of the observables to the maximum likelihood estimate of the $Z'$ mass and the $Z$--$Z'$ mixing angle as well as to the exclusion reach and statistical efficiency of the signal is investigated in details.
\end{abstract}
\maketitle

\section{Introduction}

The International Linear Collider (ILC) is discussed in the literature as a future experiment in the high energy physics \cite{Baer:2013cma}. This engine is expected to collide partially polarized electrons and positrons at the center-of-mass energies up to 1 TeV. The ILC will allow to perform precise tests of the Standard Model (SM) and beyond being a natural perspective for the current experiments on the CERN Large Hadron Collider (LHC). The combined analysis of future data from the ILC with the data obtained at the LHC is also a point of interest (for example, see Ref.  \cite{Weiglein:2004hn}).

Searches for new particles beyond the SM is one of the basic parts of the ILC experimental program. In this paper we focus on the $Z'$ boson arising in a bunch of popular models. We consider $e^+e^-\to \mu^+\mu^-$ process with the simplest annihilation kinematics at the center-of-mass energies 250 GeV, 500 GeV, and 1 TeV. Taking into account the actual bounds on the $Z'$ mass ($\sim 4$ TeV) derived from various experiments \cite{Aaboud:2017buh,Sirunyan:2018exx}, we conclude that the energy of collisions at the ILC will be significantly below the $Z'$ resonance. This means the $Z'$ boson could manifest itself through tiny contact couplings between fermionic currents induced by intermediate $Z'$ virtual states. Therefore, amplification of the corresponding signal is of great interest.

Usual observables, such as the total cross section $\sigma_{T}$ and the forward-backward asymmetry $A_{FB}$, might be essentially upgraded to increase statistical resolution as much as possible. To achieve the goal, we propose an observable constructed by integration of the differential cross section over the scattering angle with a properly chosen weight function. Such a scheme generalizes the idea of well known forward-backward or center-edge cross sections based on step-like weights for different scattering angles. In our approach, the weight function is calculated to reach the strongest $Z'$ signal with respect to the statistical noise. The corresponding integrated cross section is called {\it the optimal observable}. In a model independent approach, amplification of signals of the $Z'$ boson by means of the weighted integrated cross section was discussed in Ref. \cite{Gulov:2013kpa}.

The optimal observables are known in the high energy physics, although they are unfortunately paid no attention in searching for $Z'$ boson. They were initially applied to the analysis of the magnetic and electric dipole moments of the $t$-quark \cite{Atwood:1991ka} and to the measurement of triple gauge boson couplings at the CERN Large Electron-Positron Collider (LEP) \cite{Diehl:1993br,Diehl:2002nj}. The recent usage of the optimal observables is the investigation of the CP invariance in vector-boson fusion production of the Higgs boson at the LHC \cite{Aad:2016nal}. The optimal observables were actually rediscovered in Ref. \cite{Gulov:2013kpa}. So, the present paper could be also considered as an introduction to the optimal observables in application to the $Z'$ boson phenomenology. The resulting weight functions to integrate the differential cross section coincides with the general theory given in Refs. \cite{Atwood:1991ka,Diehl:1993br}.

We will discuss that the optimal observable is equivalent replacement of the $\chi^2$ fit of the differential cross section. This means that there is a unique weight function in the phase space to integrate the cross section {without losses of information} encoded in the differential cross section. In other words, instead of collecting events into bins and further $\chi^2$ analysis, all the events could be summed up directly with the predefined weights dependent on the scattering angle. This gives a signal of the highest quality allowed by the luminosity. So, the proposed scheme might be considered as a convenient alternative of the analysis of the differential cross section in searches for the $Z'$ boson. Moreover, it can be applied even in the case when the statistics is not rich enough to collect and publish the differential cross section.

Many models of the `new physics'  beyond the SM have been developed, and the $Z'$ boson is a usual ingredient of them. In practical searches for the $Z'$ boson, a pool of models is traditionally selected. The earlier set included the models related to different branches of the grand unification theory based on the $\mathrm{E}_6$ gauge group \cite{Hewett:1988xc,Leike:1998wr,Rizzo:2006nw,Langacker:2008yv}. Then, it was enlarged by the Alternative Left-Right Model, the Littlest Higgs Model, etc. At the moment, the LHC Collaborations discuss about ten $Z'$ models in data analysis. We consider the models discussed in the ILC Technical Design Report \cite{Baer:2013cma} and some others:
\begin{itemize}
\item In the {\em Sequential Standard Model (SSM)}, the $Z'$ couplings to fermions coincide with the SM $Z$ couplings. There is no unification of interactions in this test model. However, it is useful to clarify the definitions of $Z'$ couplings;
\item {\em $\g{E}_6$ models} \cite{Gursey:1975ki} are based on the gauge breaking scheme $\g{E}_6 \to \g{SO}(10) \times \g{U}(1)_\psi \to \g{SU}(5)\times \g{U}(1)_\chi \times \g{U}(1)_\psi \to \g{SU}(3)_C \times \g{SU}(2)_L \times \g{U}(1)_Y \times \g{U}(1)_{\theta_{E_6}}$. The model contains a free parameter -- the mixing angle $\beta$ between the $\psi$ and $\chi$ symmetry states. The angles $\beta =0$, $\pi/2$ and $\arctan(-\sqrt{5/3})$ correspond to the models called $\chi$, $\psi$ and $\eta$;
\item {\em The Left-Right Model (LR)} \cite{Pati:1974yy} is related to the gauge breaking scheme $\g{SO}(10) \to \g{SU}(3)_C \times \g{SU}(2)_L \times \g{U}(1)_Y \times \g{U}(1)_\chi \to \g{SU}(3)_C \times \g{SU}(2)_L \times \g{SU}(2)_R \times \g{U}(1)_{B-L}$. There is a free model parameter $\sqrt{2/3}\le\alpha\le\sqrt{c_W^2/s_W^2-1}$. The minimal value $\alpha=\sqrt{2/3}$ coincides with the $\chi$ model. The maximal value $\alpha=\sqrt{c_W^2/s_W^2-1}$ corresponds to the {\em Left-Right Symmetric Model (LRS)}.
\item {\em The Alternative Left-Right Model (ALR)} \cite{Ma:1986we} provides the gauge breaking scheme $\g{SO}(10) \to \g{SU}(3)_C \times \g{SU}(2)_L \times \g{U}(1)_Y \times \g{U}(1)_\chi\to \g{SU}(3)_C \times \g{SU}(2)_L \times \g{SU}(2)_R \times \g{U}(1)_{B-L}$. The ALR model is also discussed in Ref. \cite{Ashry:2013loa};
\item {\em The Littlest Higgs Model (LH)} \cite{ArkaniHamed:2002qy} considers the gauge breaking scheme $\g{SU}(5) \to [\g{SU}(2)_1 \times \g{U}(1)_1] \times [\g{SU}(2)_2 \times \g{U}(1)_2] \to \g{SU}(2)_L \times \g{U}(1)_Y$. The model is also discussed in Ref. \cite{Han:2003wu};
\item {\em The Simplest Little Higgs Model (SLH)} \cite{Kaplan:2003uc} is based on the gauge breaking scheme $\g{SU}(3)_w \times \g{U}(1)_X \to \g{SU}(2)_w \times \g{U}(1)_Y$. Two alternatives are proposed within the model: the {\em Universal SLH (USLH)} and the {\em Anomaly-Free SLH (AFSLH)}. However, both the alternatives lead to equal leptonic couplings. The model is also discussed in Ref. \cite{CortesMaldonado:2011pi}.
\item {\em The $\g{U}(1)_X$ Model} \cite{Oda:2015gna} was introduced recently as a minimal U(1) extension of the SM with conformal invariance at the classical level. We choose the $Z'$ coupling and the free model parameters in accordance with \cite{Das:2016zue}, which ensure the vacuum stability in the model.
\end{itemize}

In every $Z'$ model mentioned, the $Z'$ couplings are known, whereas the $Z'$ mass and the $Z$--$Z'$ mixing angle remain to be arbitrary parameters to be fitted in experiment. The $Z$--$Z'$ mixing angle is bounded experimentally at least as $|\sin \theta_0|<10^{-3}$ \cite{Agashe:2014kda} and usually neglected in data analysis \cite{Hewett:1988xc,Leike:1998wr,Rizzo:2006nw,Langacker:2008yv,Baer:2013cma,Osland:2009dp}. In this paper, we discuss the $Z$--$Z'$ mixing separately.

The paper is organized as follows. In section 2 we consider the differential cross section of $e^+e^-\to\mu^+\mu^-$ process and group the $Z'$ models into four different pools in dependence on the $Z'$ couplings to leptons. In section 3 the optimal observable for the $Z'$ signal is derived as an analytic solution by maximization of the signal to uncertainty ratio. In section 4 we show that the optimal observable is equivalent to the $\chi^2$ fit of the differential cross section and can be used to derive confidential intervals for the $Z'$ parameters. Effects of the $Z$--$Z'$ mixing is considered in details in Sect. 5. In Discussion section we estimate the exclusion reach for the $Z'$ mass and compare the optimal observables with the popular approach of data fitting based on the forward-backward asymmetry.

\section{The differential cross sections}

The expected scale of the grand unification as well as the $Z'$ mass is much larger than the ILC center-of-mass energies. So, the $Z'$ boson phenomenology at the ILC can be described by contact interactions between fermionic currents. We use the Lagrangian of neutral currents in the standard notations \cite{Hewett:1988xc,Leike:1998wr,Rizzo:2006nw,Langacker:2008yv,Ashry:2013loa,Han:2003wu,CortesMaldonado:2011pi}:
\begin{eqnarray}
-L_{NC}&=&e A_{\beta} J_{A,\beta} +g_Z Z_{\beta} J_{Z,\beta} +g_{Z'} Z_{\beta}^\prime J_{Z',\beta},
\nonumber\\
J_{A,\beta} &=& \sum_f \bar f \gamma^\beta  v_f^{(0)} f,
\nonumber\\
J_{Z,\beta} &=& \sum_f \bar f \gamma^\beta  (v_f-\gamma_5a_f) f,
\nonumber\\
J_{Z',\beta} &=& \sum_f \bar f \gamma^\beta  (v_f^\prime-\gamma_5a_f^\prime) f,
\label{lag}
\end{eqnarray}
where all the SM fermions $f$ appear in the sum, $A$ is the photon, $Z$ is the SM neutral vector boson, $Z'$ is the new heavy neutral vector boson, $e=\sqrt{4\pi \alpha_\mathrm{em}}$ is the positron charge, $g_Z$ and $g_{Z'}$ are the couplings to the corresponding boson (see Table \ref{Table:gZp}, $g_Z=g_{Z',\mathrm{SSM}}$), $v_f^{(0)}$ is the fermion electric charge in $e$ units, $v_f$ and $a_f$ are the vector and axial-vector coupling of the fermion to the $Z$ boson, $v'_f$ and $a'_f$ are the couplings to the $Z'$ boson. The $Z'$ boson couplings are collected in Table \ref{Table:couplings}.

\begin{table}
\caption{\label{Table:gZp} The $Z'$ coupling in the models. The cosine and sine of the Weinberg angle are denoted by $c_W$, $s_W$.}
\centering
\begin{tabular}{cc} 
\hline\hline
 & $g_{Z^\prime}$\\ 
\hline
SSM & $e/(s_W c_W)$\\ 
$\g{E}_6$, LR & $e/c_W$\\
ALR & $e/(s_W\, c_W\sqrt{1-2s_W^2})$\\
LH & $e/s_W$\\
USLH, AFSLH & $e/(c_W \sqrt{3-4s_W^2})$\\
$\g{U}(1)_X$ & $e/(4c_W)$\\
\hline\hline
\end{tabular}
\end{table} 

\begin{table}
\caption{\label{Table:couplings} The $Z'$ couplings to fermions. The sine of the Weinberg angle is denoted by $s_W$.
For the $\g{E}_6$ models, $A=\frac{\cos \beta}{2\sqrt 6}$, $B=\frac{\sqrt {10}\sin \beta }{12}$. The special values of the mixing angle $\beta =0$, $\pi/2$, and $\arctan(-\sqrt{5/3})$ correspond to $\chi$, $\psi$ and $\eta$ models. In the LR model, $\sqrt{2/3}\le\alpha_{LR}\le\sqrt{c_W^2/s_W^2-1}$. The LRS model corresponds to $\alpha=\sqrt{c_W^2/s_W^2-1}$. The LH model includes the parameter $\frac{1}{10}\le \frac{c}{s}\le 2$, we choose $\frac{c}{s}\equiv 1$ \cite{Han:2003wu}. In the $\g{U}(1)_X$, we choose $x_\Phi=2$ and $x=x_H/x_\Phi=1$ or $-1.25$.}
\centering
\begin{tabular}{ccccc} 
\hline\hline
$f$ & $\nu$ & $e$ & $u$ & $d$ \\ 
\hline\multicolumn{5}{c}{\rule[-5pt]{0pt}{18pt}SSM}\\
$2v_f^\prime$ & $\frac{1}{2}$ & $2s_W^2-\frac{1}{2}$ & $\frac{1}{2}-\frac{4}{3} s_W^2$ & $\frac{2}{3} s_W^2-\frac{1}{2}$ \\[3pt]
$2a_f^\prime$ & $\frac{1}{2}$ & $-\frac{1}{2}$ & $\frac{1}{2}$ & $-\frac{1}{2}$ \\[3pt]
\hline\multicolumn{5}{c}{\rule[-5pt]{0pt}{18pt}$\g{E}_6$}\\ 
$2v_f^\prime$ & $3A+B$ & $4A$ & $0$ & $-4A$ \\[3pt]
$2a_f^\prime$ & $3A+B$ & $2(A+B)$ & $2(B-A)$ & $2(A+B)$ \\[3pt]
\hline\multicolumn{5}{c}{\rule[-5pt]{0pt}{18pt}LR}\\
$2v_f^\prime$ & ${\frac{1}{2\alpha}}$ & ${\frac{1}{\alpha}}- {\frac{\alpha}{2}}$ & ${\frac{\alpha}{2}}-{\frac{1}{ 3\alpha}}$ & $-{\frac{1}{3\alpha}}-{\frac{\alpha}{2}}$ \\[3pt]
$2a_f^\prime$ & ${\frac{1}{2\alpha}}$ & ${\frac{\alpha}{2}}$ & $-{\frac{\alpha}{2}}$ & ${\frac{\alpha}{2}}$ \\[3pt]
\hline\multicolumn{5}{c}{\rule[-5pt]{0pt}{18pt}ALR}\\
$2v_f^\prime$ & ${s_W^2-\frac{1}{2}}$ & ${\frac{5}{2}s_W^2-1}$ & ${\frac{1}{2}-\frac{4}{3}s_W^2}$ & ${\frac{1}{6}s_W^2}$ \\[3pt]
$2a_f^\prime$ & ${s_W^2-\frac{1}{2}}$ & ${-\frac{1}{2}s_W^2}$ & ${s_W^2-\frac{1}{ 2}}$ & $-\frac{1}{2}s_W^2$ \\[3pt]
\hline\multicolumn{5}{c}{\rule[-5pt]{0pt}{18pt}LH}\\
$2v_f^\prime$ & $\frac{c}{4s}$ & $-\frac{c}{4s}$ & $\frac{c}{4s}$ & $-\frac{c}{4s}$ \\[3pt]
$2a_f^\prime$ & $\frac{c}{4s}$ & $-\frac{c}{4s}$ & $\frac{c}{4s}$ & $-\frac{c}{4s}$ \\[3pt]
\hline\multicolumn{5}{c}{\rule[-5pt]{0pt}{18pt}USLH}\\
$2v_f^\prime$ & $\frac{1}{2}-s_W^2$ & $\frac{1}{2}-2s_W^2$ & $\frac{1}{2}+\frac{1}{3} s_W^2$ & $\frac{1}{2}-\frac{2}{3} s_W^2$ \\[3pt]
$2a_f^\prime$ & $\frac{1}{2}-s_W^2$ & $\frac{1}{2}$ & $\frac{1}{2}-s_W^2$ & $\frac{1}{2}$ \\[3pt]
\hline\multicolumn{5}{c}{\rule[-5pt]{0pt}{18pt}AFSLH}\\
$2v_f^\prime$ & $\frac{1}{2}-s_W^2$ & $\frac{1}{2}-2s_W^2$ & $-\frac{1}{2}+\frac{4}{3} s_W^2$ & $\frac{1}{3} s_W^2-\frac{1}{2}$ \\[3pt]
$2a_f^\prime$ & $\frac{1}{2}-s_W^2$ & $\frac{1}{2}$ & $-\frac{1}{2}$ & $s_W^2-\frac{1}{2}$ \\[3pt]
\hline\multicolumn{5}{c}{\rule[-5pt]{0pt}{18pt}$\g{U}(1)_X$}\\
$2v_f^\prime$ & $-x_H - x_\Phi$ & $-3x_H - x_\Phi$ & $\frac{5}{3}x_H + \frac{1}{3}x_\Phi$ & $-\frac{1}{3}x_H + \frac{1}{3}x_\Phi$ \\[3pt]
$2a_f^\prime$ & $-x_H$ & $x_H$ & $-x_H$ & $x_H$ \\[3pt]
\hline\hline
\end{tabular}
\end{table} 

The differential cross section of the process $e^+e^-\to \mu^+\mu^-$ consists of the SM part and the $Z'$ contribution. Since we assume the energies significantly below the threshold of $Z'$ decoupling, $\sqrt{s}\ll M_{Z'}$, it can be expanded by a small parameter:
\begin{eqnarray}
&&\frac{d\sigma}{dz}= \frac{d\sigma^\mathrm{SM}}{dz} + \sum\limits_{n=1}^{\infty}\mu^n\, F_n(s,z,a',v'),
\label{otdif}
\\
&&\mu=\frac{M^2_Z}{s - M^2_{Z'}},
\label{mu}
\end{eqnarray}
where $z$ is the cosine of the scattering angle of the charged lepton in the center-of-mass frame, $M$ means the mass of the corresponding particle, $F$ are factors measured in the same units as the cross section, $\sqrt{s}$ is the center-of-mass energy. 
The magnitude of the expansion parameter $\mu$ is about $10^{-4}$ for $\sqrt{s}= 1$ TeV and $M_{Z'}=4$ TeV. Heavier $Z'$ masses and lower collision energies give smaller $\mu$. As it is seen, the $Z'$ mass, measured in units of $M_Z$, plays the role of unknown dimensionless parameter of the model, whereas other components of (\ref{otdif}) can be calculated numerically. We neglect the widths of vector bosons, since they are of few percents of the boson mass in the considered models and the energies are aside from the $Z$ and $Z'$ peaks.

\begin{figure}
\includegraphics[width=0.45\textwidth]{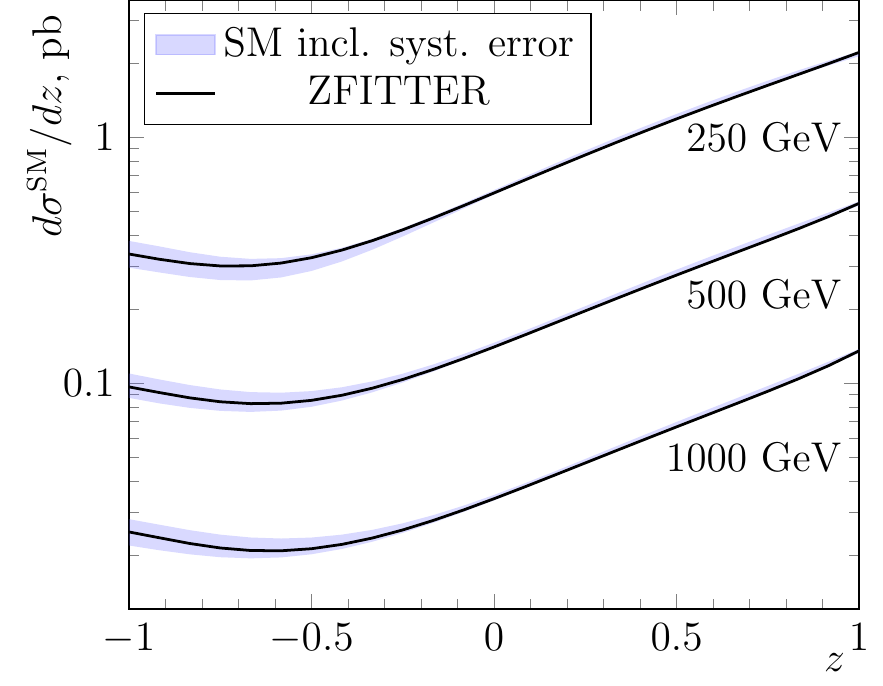}
\caption{The SM differential cross section (unpolarized) in pb with 2\% systematical error used in numeric calculations. The results of ZFITTER are plotted as lines.}\label{fig:dsm}
\end{figure}

In our paper, the SM differential cross section is calculated by two complementary approaches. First, we use FeynArts \cite{Hahn:2000kx}, FormCalc \cite{Hahn:1998yk} and LoopTools \cite{Hahn:1998yk} up to one-loop radiative corrections for the weak sector. Effects of the quantum electrodynamics is taken into account in accordance with \cite{Bardin:1999ak}: the soft photon bremsstrahlung is included analytically, whereas the hard photon bremsstrahlung included by numerical integration in the phase space of the final state. The domain in the phase space is determined by the event selection rule $\sqrt{{s'}/{s}}>0.85$, where $s'$ is the Mandelstam variable of the final pair $\mu^+\mu^-$, and $s$ is the Mandelstam variable of the final state with the photon $\mu^+\mu^-\gamma$. Second, the SM differential cross section is computed by ZFITTER software \cite{Arbuzov:2005ma}. The discrepancy between the results is less than 2\%. So, we add 2\% systematic error to the SM differential cross section obtained in the first approach (Fig. \ref{fig:dsm}). This systematic error covers also four-fermion final states with leptons missed in the beams, which contribution is estimated to be less than 1\% \cite{Gulov:2013kpa}.

The leading-order (LO) $Z'$ factor $F_1$ in (\ref{otdif}) arises from the interference between the SM amplitude and the $Z'$ exchange amplitude. For our purposes it is calculated in the improved Born approximation with the running constants. The factor can be written in the following form:
\begin{eqnarray}
&&F_1(s,z,a',v')= \frac{\alpha_\mathrm{em}g_{Z^\prime}^2}{32\sin^2\theta_W\cos^2\theta_W M^2_Z}
\nonumber\\&&
\quad\times
\left[z\,f(s,a'_e,v'_e) + \frac{1+z^2}{2}f(s,v'_e,a'_e)\right],
\label{F}
\end{eqnarray}
with
\begin{equation}
f(s,x,y) = x^2 (1 - \epsilon) (3 + \epsilon) + (y + \epsilon\,x)^2  \frac{s}{s-M^2_Z},
\label{f}
\end{equation}
where $\theta_W$ is the Weinberg angle, and $\epsilon=1-4\sin^2\theta_W$. The small parameter $\epsilon$ is less than 2\%, so it can be set to zero in qualitative analysis. Since both the $Z'$ couplings $a'$ and $v'$ are of the same order, the contribution of the first term in $f$ is approximately 2.5-3 times larger than the contribution from the second term (at the ILC energies). So, the second argument in $f$ dominates over the third argument. This means that
\begin{itemize}
\item
the $z$-odd part of $F_1$ is mainly related to the axial-vector $Z'$ coupling $a'_e$;
\item
the $z$-even part of $F_1$ is mainly related to the vector $Z'$ coupling $v'_e$;
\item
the factor $F_1$ weakly depends on the collision energy.
\end{itemize}

\begin{figure}
\includegraphics[width=0.45\textwidth]{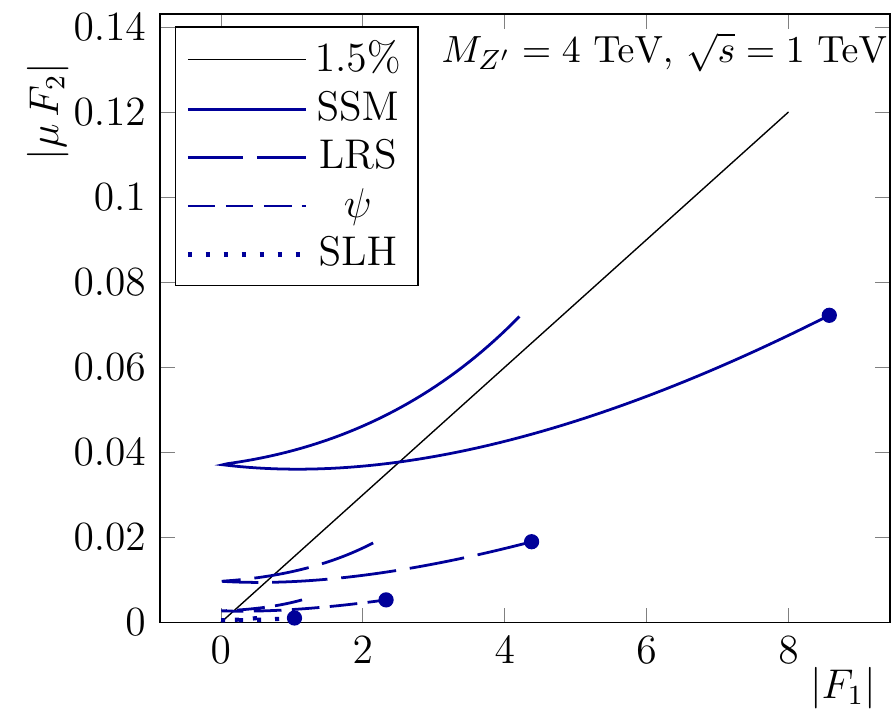}
\includegraphics[width=0.45\textwidth]{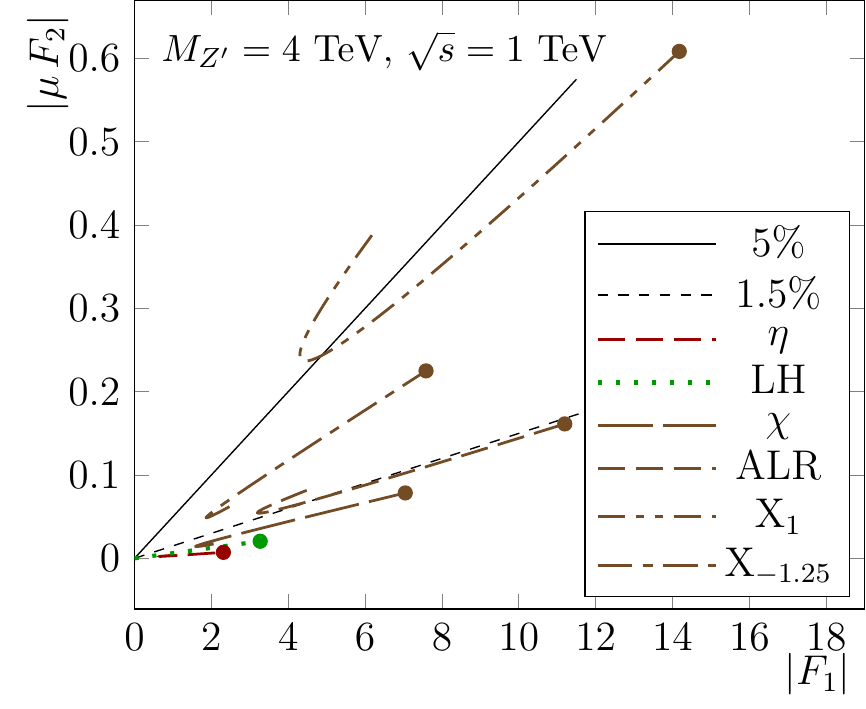}
\caption{
Estimation of the NLO contribution in $\mu$ to the cross section. The $Z'$ mass and the collision energy correspond to the maximal influence from $F_2$, $\mu=-6.7\times 10^{-4}$. The dots show the maximal values of $F_1$ used to set the level of systematic errors. In $\g{U}(1)_{X,x=-1.25}$ and $\g{U}(1)_{X,x=-1.25}$ models (marked as $\mathrm{X}_{-1.25}$ and $\mathrm{X}_{1}$), the NLO term is below 4\% of $F_1$. For other models, this term is below 1.5\%. Lower energies and higher $M_{Z'}$ suppress the relative contribution from $F_2$. The plotted factors for the $\g{U}(1)_X$ model must be multiplied by 2.5.
}\label{fig:fsyst}
\end{figure}

The next-to-leading-order (NLO) factor $F_2$ in (\ref{otdif}) is mainly determined by the squared amplitude with an intermediate $Z'$ state. Its improved Born approximation is
\begin{eqnarray}
&&F_2(s,z,a',v')= \frac{g_{Z^\prime}^4 s}{32\pi M^4_Z}
\nonumber\\&&\quad\times
\left[8a^{\prime 2}_e v^{\prime 2}_e z +\left(a^{\prime 2}_e + v^{\prime 2}_e\right)^2 (1+z^2)\right].
\label{F2}
\end{eqnarray}
In order to estimate the NLO contribution to the cross section, we compare $F_1$ and $\mu\,F_2$ at the highest ILC energy (1 TeV) and the lowest $Z'$ mass (4 TeV). The corresponding value of the expansion parameter $\mu=-6.7\times 10^{-4}$. These settings give the maximal possible contribution beyond the LO. The comparison between $F_1$ and $\mu\,F_2$ is shown in Fig. \ref{fig:fsyst}, where the lines cover different values of $z$. The dots show the maximal values of $F_1$ (at $z=1$) used to set the level of systematic errors. In the $\g{U}(1)_{X,x=-1.25}$ and $\g{U}(1)_{X,x=-1.25}$ models, the NLO term is below 4\% of the leading term. For other models, it is below 1.5\%. Of course, energies below 1 TeV and $M_{Z'}>4$ TeV give smaller relative contributions from $F_2$.

\begin{figure}
\includegraphics[width=0.45\textwidth]{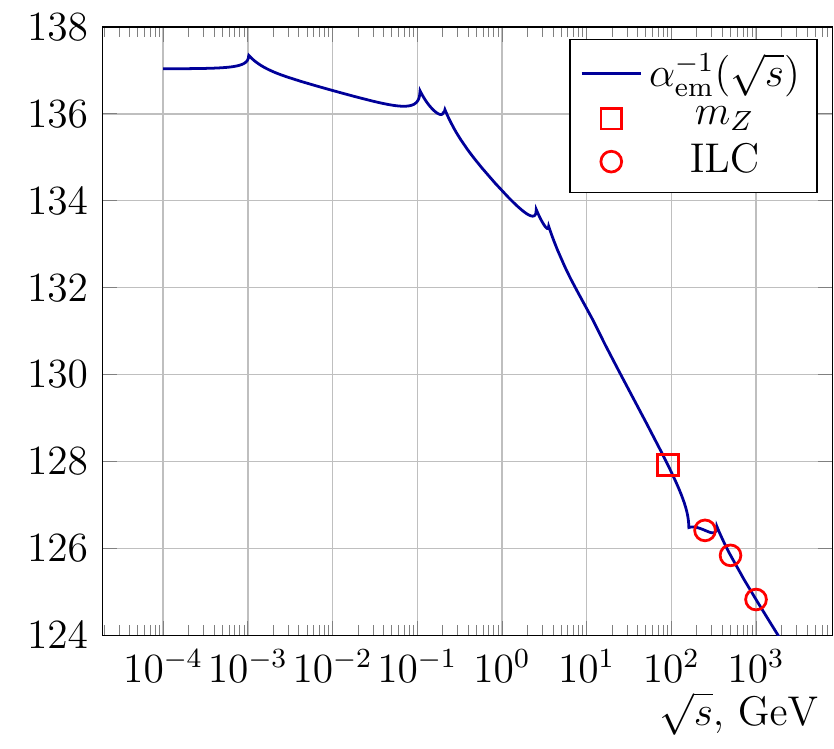}
\caption{The fine structure constant $\alpha_\mathrm{em}$ determined by the photon polarization operator in a wide interval of energies. The values at $m_Z$ as well as at the supposed ILC energies are shown separately.}\label{fig:alpha}
\end{figure}

The fine structure constant $\alpha_\mathrm{em}$ is determined by the photon polarization operator (see Fig. \ref{fig:alpha}). The Weinberg angle is taken in accordance with \cite{Erler:2004in}, so it is higher than the value at the $Z$ peak used, for example, in \cite{Han:2013mra}. The numeric values can be found in Table \ref{Table:runningcouplings}. As it is seen from (\ref{F}), the systematic error from the fine structure constant is factorized and irrelevant for the angular behavior of the $Z'$ factor. On the other hand, the systematic error of the Weinberg angle has to be taken into account during angular integrations. We can estimate it by means of the first (leading) term in (\ref{f}):
\begin{equation}
\left|\frac{\delta_\mathrm{syst.}F_1}{F_1}\right|\simeq
\left|\frac{\delta_\mathrm{syst.}f}{f}\right|\simeq
0.7|\delta_\mathrm{syst.}\epsilon|.
\label{Fsyst}
\end{equation}
Comparing the values in Table \ref{Table:runningcouplings} with the effective leptonic Weinberg angle at the $Z$ peak ($\simeq 0.23$), we conclude that the non-factorizable radiative corrections in $F_1$ is up to 4\%. Thus, all the systematic errors (including radiative corrections and the expansion in $\mu$) are estimated below 4\%. In further calculations, we will use the LO factor $F_1$ and add the systematic error of order 5\% of the maximal value of $F_1(z)$ to take into account effects beyond the approximation used.

\begin{table}
\caption{\label{Table:runningcouplings} The running couplings used to calculate the $Z'$ contribution to the cross section.}
\centering
\begin{tabular}{cccc} 
\hline\hline
 & 250 GeV & 500 GeV & 1 TeV \\ 
\hline
$1/\alpha_\mathrm{em}$ & 126.414 & 125.839 & 124.823 \\[3pt]
$\sin^2\theta_W$ & 0.2368 & 0.2407 & 0.2446 \\[3pt]
\hline\hline
\end{tabular}
\end{table} 

\begin{figure}
\includegraphics[width=0.45\textwidth]{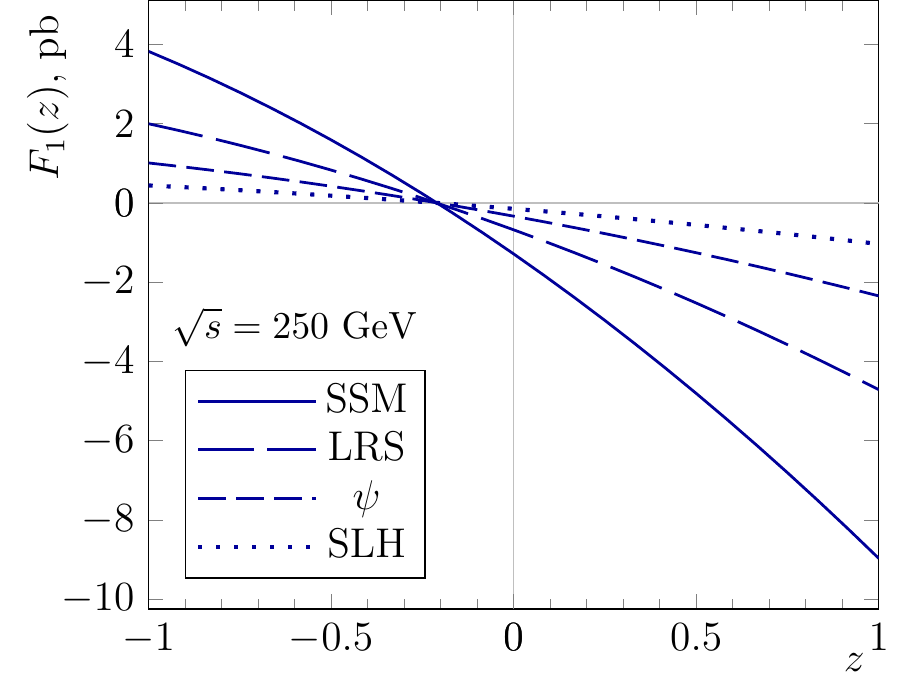}
\includegraphics[width=0.45\textwidth]{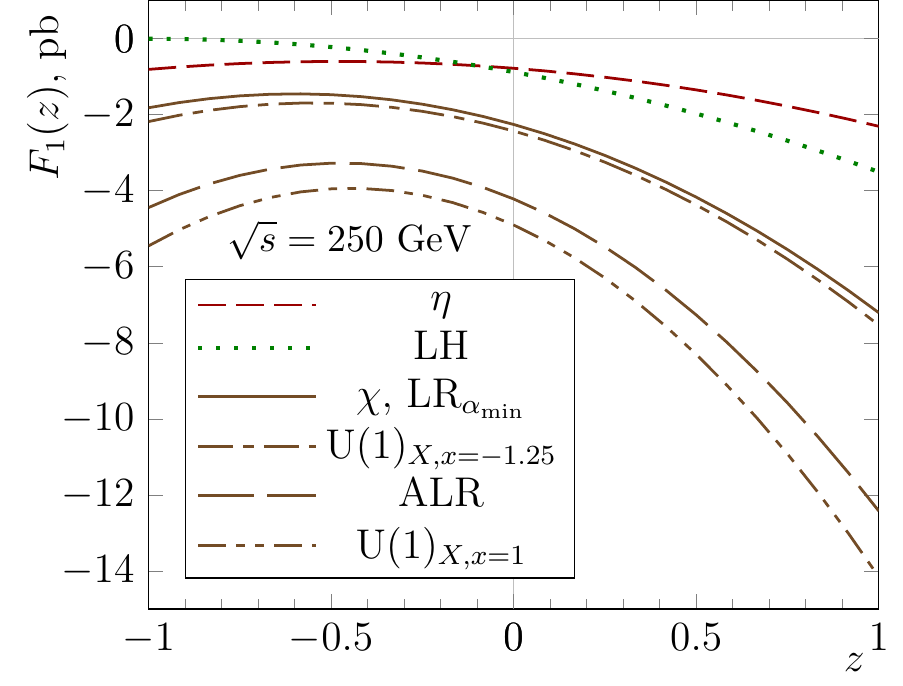}
\caption{The unpolarized factors $F_1(\sqrt{s},z)$ in pb for models at 250 GeV. The factors are calculated up to 5\% systematic error. The plotted factors for the $\g{U}(1)_X$ model must be multiplied by 2.5.}\label{fig:trrk250}
\end{figure}

The $Z'$ factors in the models are shown in Fig. \ref{fig:trrk250} for the center-of-mass energy 250 GeV only, since there is no dramatic changes in their shapes for higher energies. We can select four different groups of models:
\begin{itemize}
\item {\em SSM, $\psi$, LRS and SLH}. In these models, {\it the $Z'$ vector coupling to charged leptons is suppressed}. This could be seen by substitution $s_W^2\simeq 1/4$, $A=0$, $\alpha\simeq\sqrt{2}$ in Table \ref{Table:couplings}. The $\g{U}(1)_X$ model with $x_H/x_\Phi=-1/3$ also belongs to this class. Thus, the $Z'$ factors depend mainly on the axial-vector coupling and, consequently, on the $z$-odd term in (\ref{f}). The factors are approximately proportional and odd with respect to $z$, so one could naively expect that the best signal would be described by the forward-backward cross section.
\item {$\eta$}. {\it The suppressed $Z'$ axial-vector coupling to charged leptons} is a feature of the model. The axial-vector coupling vanishes exactly at $A=-B$ in Table \ref{Table:couplings} ($\beta=\arctan(-\sqrt{3/5})\simeq -0.21\pi$). For the $\eta$ model, $\beta=\arctan(-\sqrt{5/3})\simeq -0.29\pi$ and $A\simeq-B$. The $Z'$ factor depends mainly on the vector coupling and, consequently, on the $z$-even term in (\ref{f}). One could naively expect that the best signal would be described by the total cross section.
\item {\em LH}. The vector and the axial-vector couplings to charged leptons are equal. This means {\it interactions with the left-handed chiral states only}. The $\g{E}_6$ model with $A=B$ ($\beta=\arctan\sqrt{3/5}$, the so called $I$ model), the LR model with $\alpha=1$, and the $\g{U}(1)_X$ model with $x_H/x_\Phi=-1/4$ also belong to this class. The angular dependence in this case is $F\sim (1+z)^2$, so the factor contains dominant contribution from the forward scattering angles. One could naively expect that the best signal would be described by the forward cross section.
\item {\em $\chi$, $\mathrm{LR}_{\alpha_\mathrm{min}}$, $\g{U}(1)_{X,x=-1.25}$, ALR, $\g{U}(1)_{X,x=1}$}. The models with mixed angular dependence of the $Z'$ factor. Let us note that the first two models coincide exactly.
\end{itemize}

As we will see in the next section, naive expectations about observables to amplify $Z'$ signals do not correspond to the best choice. This is because the statistical error is not uniform over the scattering angle and has to be also taken into consideration to find the strongest signal of the particle.

\section{The optimal observable}\label{secOO}

The optimal observable to select $Z'$ signal is defined by weighted integration of the cross section:
\begin{equation}
\sig=\int_\Omega dz\ w(z)\ \frac{d\sigma}{dz},
\end{equation}
where $w(z)$ is the weight function and $\Omega$ is the interval of available scattering angles. The complete phase space is given by $z\in [-1,1]$. We will specify $\Omega$ when it influences the result.

The standard deviation of the observable can be calculated from the Poisson distribution of events (see \cite{Gulov:2013kpa}):
\begin{equation}
\delta\sig\simeq\sqrt{\frac{1}{{\cal L}_\mathrm{eff}} \int_\Omega dz~ w^2(z)\ \frac{d\sigma^\mathrm{SM}}{dz}}.
\end{equation}
where ${\cal L}$ is the luminosity, and the cross section is substituted by its SM part, since the $Z'$ contribution is tiny and leads to higher order corrections in the inverse $Z'$ mass. The SM cross section can be redefined to take into account the acceptance rate of events. We introduce the `effective' luminosity corrected by the polarization of the input beams: ${\cal L}_\mathrm{eff}=(1+P^+P^-){\cal L}$. In numeric estimates the following integrated luminosities are assumed: ${\cal L}_\mathrm{250\ GeV}=250\mbox{ fb}^{-1}$, ${\cal L}_\mathrm{500\ GeV}=500\mbox{ fb}^{-1}$, and ${\cal L}_\mathrm{1\ TeV}=1000\mbox{ fb}^{-1}$ \cite{Baer:2013cma}. The polarizations of the initial electron and positron states are $P^-=\eta_{e^-_L}-\eta_{e^-_R}$ and $P^+=\eta_{e^+_R}-\eta_{e^+_L}$, where $\eta$ is the fraction of the corresponding particles. The definition of polarization is in accordance with \cite{Han:2013mra}. 

The optimal observable satisfies the condition
\begin{equation}
\frac{\mathrm{abs}(\mbox{signal})}{\mbox{uncertainty}}
=\frac{\mathrm{abs}(\sig-\sig^\mathrm{SM})}{\delta\sig}
\to \mathrm{max},
\label{otnoso}
\end{equation}
where the $Z'$ signal is the deviation from the SM, i.e. the SM is considered as the background. The uncertainty is the statistical error (the standard deviation of the observable). This condition is actually a functional of the weight function $w$. Its maximum in the Hilbert space of $w$ determines uniquely the weight function and the optimal observable.

In the considered $Z'$ models, the theoretical prediction of the signal is
\begin{equation}
\sig-\sig^\mathrm{SM}
\simeq\mu \int_\Omega dz\ w(z)\ F_1(z).
\label{nablud}
\end{equation}
As it is seen, the unknown parameter factorizes, and the weight function is {\it independent of $\mu$}:
\begin{equation}
\mathrm{abs}\left[\frac{\displaystyle \int_\Omega dz\ w(z)\ F_1(z)}
{\sqrt{\displaystyle\int_\Omega dz~ w^2(z)\ \frac{d\sigma^\mathrm{SM}}{dz}}}\right]
\to \mathrm{max}.
\label{target}
\end{equation}

The target functional (\ref{target}) is quite simple, and the exact analytic solution can be written. Since the SM cross section is strictly positive, let us define the following scalar product in the Hilbert space:
\begin{equation}
\langle f_1,f_2 \rangle = \int_\Omega dz\ f_1(z)f_2(z)\ \frac{d\sigma^\mathrm{SM}}{dz}.
\label{sprod}
\end{equation}
Then, the denominator of  (\ref{target}) is just the norm of the weight function, $\|w\|$, whereas the numerator is the scalar product between $w$ and the function
\begin{equation}
\tilde{F}(z) = \frac{F_1(z)}{d\sigma^\mathrm{SM}/dz}.
\end{equation}
The function $\tilde{F}$ does not depend on $w$. Dividing (\ref{target}) by $\|\tilde{F}\|$, we obtain
\begin{equation}
\mathrm{abs}\left[\frac{\langle w,\tilde{F} \rangle}
{\|w\|\cdot\|\tilde{F}\|}\right]
\to \mathrm{max}.
\label{opt1}
\end{equation}
Thus, the target functional is `the cosine between vectors $w$ and $\tilde{F}$', and we immediately write the solution (up to a normalization factor $C$):
\begin{equation}
w(z) = C\,\tilde{F}(z) = C\,\frac{F_1(z)}{d\sigma^\mathrm{SM}/dz}
\label{wfexact}
\end{equation}
This solution reproduces exactly the general theory of the optimal observables \cite{Atwood:1991ka,Diehl:1993br,Diehl:2002nj}. We have additionally confirmed the result by numeric optimization described in details in Ref. \cite{Gulov:2013kpa}, but the numeric approach contains no interesting details concerning the subject of the present paper to discuss it here. We have also cross-checked that the exact solution (\ref{wfexact}) describes the numeric results obtained in Ref. \cite{Gulov:2013kpa}.

\begin{figure}
\includegraphics[width=0.45\textwidth]{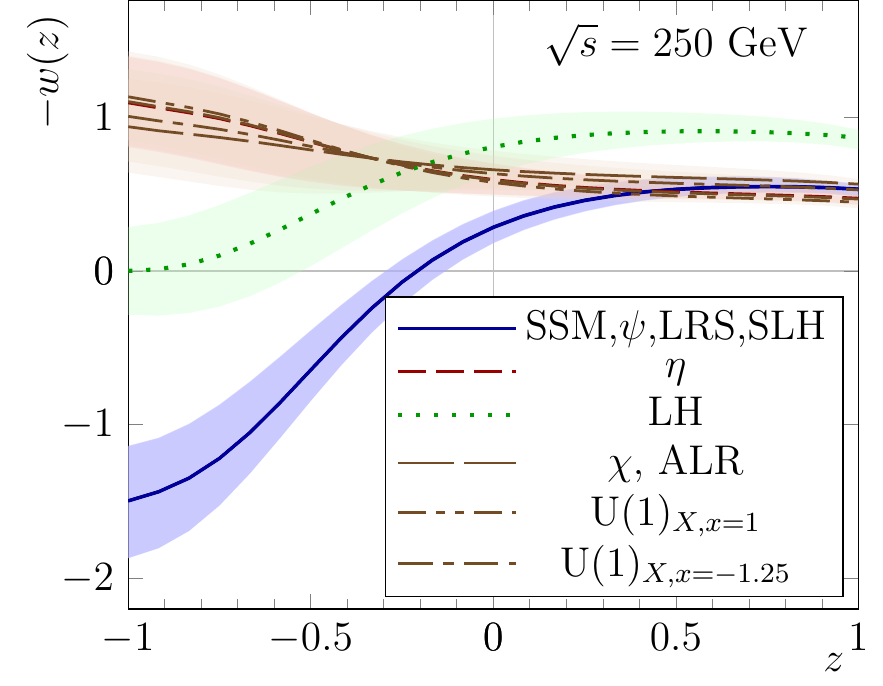}
\includegraphics[width=0.45\textwidth]{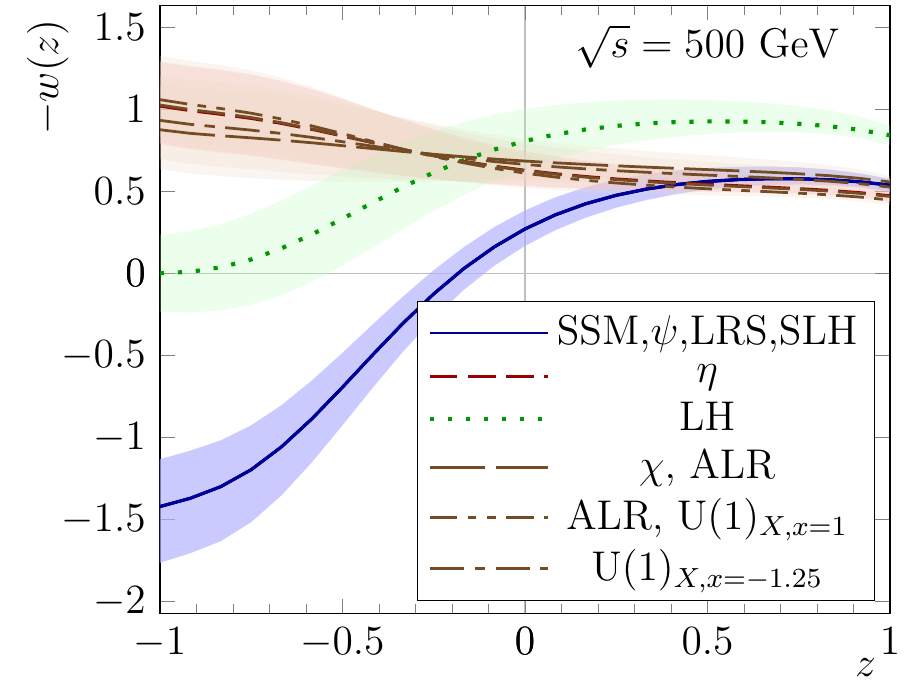}
\includegraphics[width=0.45\textwidth]{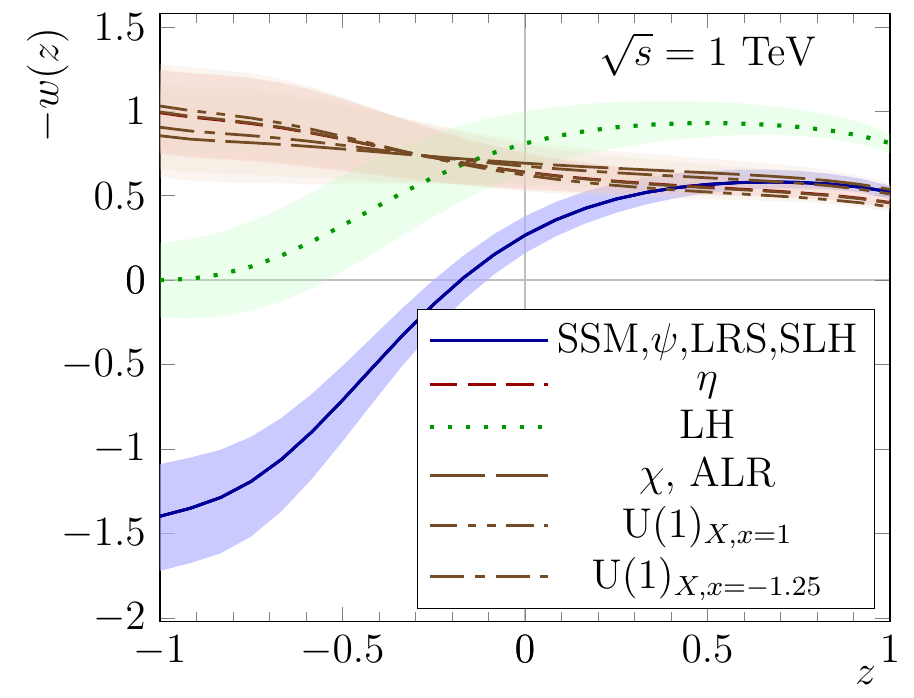}
\caption{The weight functions $-w(\sqrt{s},z)$ to amplify the $Z'$ signal and to measure $M_{Z'}$ at the ILC energies. The uncertainty arises from the systematic errors on the SM cross section and $Z'$ factors. The universal normalization $\int_{-1}^1 w^2 dz=1$ is used to compare different models. The negative sign at $w$ is chosen to obtain positive values at $z=1$. The actual systematic error might be significantly less accounting for possible destructive interference between the weight function and the uncertainties of the SM and $Z'$ contribution to the cross section. The lines correspond to the SM from ZFITTER.}
\label{fig:w}
\end{figure}

In order to compare different $Z'$ models, it is convenient to set the universal normalization of the weight function:
\begin{equation}
\int\limits_{-1}^1 w^2 dz= 1,\qquad C = \frac{1}{\displaystyle\sqrt{\int_{-1}^1 \left[
\frac{F_1(z)}{d\sigma^\mathrm{SM}/dz}
\right]^2 dz}}.
\label{wfuniv}
\end{equation}
The universally normalized weight functions are plotted in Fig. \ref{fig:w}. As it is seen, the weight functions are stable with respect to systematic errors and weakly depend on the center-of-mass energy. In computation of systematic errors, possible destructive interference between the weight function and the uncertainties of the SM and $Z'$ factor is not taken into account. So the actual systematic error might be less up to several times. Of course, more accurate calculation of the cross section could also reduce the error.

Some remnants of the `naive observables' for the $Z'$ signal can be still found in the weight functions. For the SSM-like pool of models, we see the smoothed forward-backward cross section, but the weight function is not the step function anymore. In the LH model, the observable selects mainly the forward bins ($z\ge -0.5$). Other models are closer to the total cross section with increased weight of backward scattering angles.

\section{Relation to the $\chi^2$ fit of the differential cross section}

Let us consider the fit of the $Z'$ mass from the differential cross section. The observed events are aggregated into bins. Each bin is described by the following quantities:
\begin{description}
\item[$\Delta z_{i}$] the width of the $i$th bin;
\item[$\sigma_{i}, \sigma^\mathrm{SM}_{i}$] the observed cross section integrated in the bin and the correspondent SM value:
\[
\sigma_{i} \simeq \left(\frac{d\sigma}{dz}\right)_{z_{i}} \Delta z_i;
\]
\item[$\delta_i$] the statistical error in the bin. In accordance with the Poisson distribution of events,
\[
\delta^2_i \simeq {\cal L}^{-1}\sigma_i \simeq {\cal L}^{-1}\sigma^\mathrm{SM}_i;
\]
\item[$F_{1,i}$] the $Z'$ factor integrated in the bin:
\[
F_{1,i} \simeq F_1(z_i) \Delta z_i.
\]
\end{description}

The $\chi^2$ function is
\begin{equation}
\chi^2(\mu)
=\sum_i\left(\frac{\sigma_i - \sigma^\mathrm{SM}_i -\mu F_{1,i}}{\delta_i}\right)^2
\end{equation}
The minimum of $\chi^2$ can be found explicitly. It gives the maximum likelihood (ML) estimate of the parameter:
\begin{equation}
\mu_\mathrm{ML}=
\frac{\displaystyle\sum_i\frac{(\sigma_i-\sigma^\mathrm{SM}_i)  F_{1,i}}{\delta_i^2}}{\displaystyle\sum_i\frac{F_{1,i}^2}{\delta_i^2}}.
\end{equation}
In the continuous limit,
\begin{eqnarray}
&&\mu_\mathrm{ML} = \int_\Omega dz\,w_\mathrm{ML}(z)\left(\frac{d\sigma}{dz}-\frac{d\sigma^\mathrm{SM}}{dz}\right),
\nonumber\\
&&w_\mathrm{ML}(z)=\frac{\displaystyle\frac{F_1(z)}{d\sigma^\mathrm{SM}/dz}}{\displaystyle\int_\Omega dz\ \frac{F_1^2(z)}{d\sigma^\mathrm{SM}/dz}}.
\label{MLobserv}
\end{eqnarray}
As it is seen, the ML estimate is described by the same weight function as the optimal observable described in the previous section. The only difference is the normalization of $w(z)$, which is not arbitrary in this case. The normalization condition can be written as follows:
\begin{equation}
\int_\Omega dz\ w_\mathrm{ML}(z)F_1(z) = 1.
\label{MLnorm}
\end{equation}
Thus, the weight function in Fig. \ref{fig:w} must be re-scaled to be used as the ML estimator of the $Z'$ mass. The corresponding conversion constants are summarized in Table \ref{tab:wML}. For realistic detectors, some cuts occur for the scattering angle near the beam direction. In this case, the conversion constant is larger, since the weight function is distributed over smaller volume. Note that the shape of the weight function (up to the normalization factor) does not depend on the cuts in accordance with (\ref{wfexact}).

\begin{table}
\caption{\label{tab:wML}
The conversion constants in $\mathrm{pb}^{-1}$ to obtain weight functions $w_\mathrm{ML}$ for the ML estimator (\ref{MLobserv}), (\ref{MLobservEvents}) from the weight functions normalized as $\int_{-1}^1 w^2 dz=1$ (plotted in Fig. \ref{fig:w}). The SM from ZFITTER is assumed. The complete phase space and realistic kinematic cuts are considered.
}
\begin{center}
\begin{tabular}{cccc} 
\hline\hline
 & 250 GeV &  500 GeV & 1 TeV \\ 
\hline
\multicolumn{4}{c}{$z\in[-1,1]$} \\
\hline
SSM&0.229335&0.223967&0.222132\\
$\chi$&0.255427&0.251687&0.24824\\
$\psi$&0.876493&0.839409&0.817898\\
$\eta$&0.782297&0.750646&0.729826\\
LRS&0.436601&0.432227&0.434998\\
ALR&0.144713&0.146662&0.150483\\
LH&0.5007&0.524721&0.535442\\
SLH&1.9879&1.89537&1.83567 \\
$\g{U}(1)_{X,x=1}$&0.0502789&0.0481229&0.0467804\\
$\g{U}(1)_{X,x=-1.25}$&0.0972673&0.0938825&0.0913626\\
\hline
\multicolumn{4}{c}{$z\in[-0.9848,0.9848]$, ($10^\circ < \theta < 170^\circ$)} \\
\hline
SSM&0.237948&0.232167&0.230131\\
$\chi$&0.261228&0.257191&0.253523\\
$\psi$&0.909494&0.870178&0.847366\\
$\eta$&0.800892&0.767772&0.746045\\
LRS&0.452973&0.448038&0.450659\\
ALR&0.148165&0.150018&0.153834\\
LH&0.512453&0.536649&0.547246\\
SLH&2.06256&1.96477&1.90178\\
$\g{U}(1)_{X,x=1}$&0.0514954&0.0492402&0.0478394\\
$\g{U}(1)_{X,x=-1.25}$&0.0995117&0.0959634&0.0933323\\
\hline\hline
\end{tabular}
\end{center}
\end{table}

The $\chi^2$ function allows also to derive the confidence interval for the $Z'$ mass. It is given by
\begin{equation}
\chi^2(\mu) - \chi^2_\mathrm{min} < \chi^2_{1,\mathrm{CL}} = N_\mathrm{CL}^2,
\label{CLdef}
\end{equation}
where $N_\mathrm{CL}$ is the symmetric confidence level (CL) for the standard normal distribution: the interval 
$[-N_\mathrm{CL},N_\mathrm{CL}]$ corresponds to probability $p_\mathrm{CL}$. In other words, $N_\mathrm{CL}$ is the CL measured in units of the standard deviation (in the so called `sigmas'). The $\chi^2$ distribution with one degree of freedom is used, since a single linear parameter appears in the fit.

Calculating (\ref{CLdef}) explicitly, we obtain the confidence interval for $\mu$:
\begin{equation}
(\mu - \mu_\mathrm{ML})^2 \sum_i\frac{F_{1,i}^2}{\delta_i^2} < N_\mathrm{CL}^2.
\label{CLN}
\end{equation}

Taking one standard deviation, $N_\mathrm{CL}=1$, we derive the statistical error of the $Z'$ signal:
\begin{equation}
\delta_\mu = \mathrm{abs}(\mu_{N_\mathrm{CL}=1} - \mu_\mathrm{ML}) = 1 \Big/ \sqrt{\sum_i\frac{F_{1,i}^2}{\delta_i^2}}.
\end{equation}

We can also calculate the quality of the $Z'$ signal. The signal means that $\mu=0$ is excluded at some CL. Taking $\mu=0$ in (\ref{CLN}), we obtain
\begin{equation}
N_\mathrm{CL,signal} =  |\mu_\mathrm{ML}| \sqrt{\sum_i\frac{F_{1,i}^2}{\delta\sigma_i^2}} = \frac{|\mu_\mathrm{ML}|}{\delta_\mu}
\label{Nsignal}
\end{equation}
or, in the continuous limit,
\begin{equation}
N_\mathrm{CL,signal} = |\mu_\mathrm{ML}|\sqrt{{\cal L}\int_\Omega dz\ \frac{F_1^2(z)}{d\sigma^\mathrm{SM}/dz}}.
\label{NsignalC}
\end{equation}

The right hand side of (\ref{Nsignal}) is just the signal-to-uncertainty ratio (\ref{otnoso}). In accordance with the previous section, it is maximal for the considered fit. Thus, the $\chi^2$ fit of the differential cross section coincides again with the optimal observable, since both the approaches are actually the same ML estimate. It is well known that the ML fit is a sort of estimates with the best statistical efficiency.

The expected signal-to-uncertainty ratio $N_\mathrm{CL,signal}$ is shown in Table \ref{Table:SilsAll} for $M_{Z'}=4$ TeV and the ILC canonical luminosities (250 fb${}^{-1}$, 500 fb${}^{-1}$, and 1000 fb${}^{-1}$ for 0.25, 0.5, and 1 TeV). It is easy obtain this ratio for other $Z'$ masses and luminosities, since $N_\mathrm{CL,signal}\sim \sqrt{\cal L}/(s-M^{2}_{Z'})$. The condition $N_\mathrm{CL,signal}> 5$ means the discovery of the particle.

\begin{table}
\caption{\label{Table:SilsAll} The expected signal-to-uncertainty ratio $N_\mathrm{CL,signal} =  \frac{\sig-\sig^\mathrm{SM}}{\delta\sig}$ corresponding to the optimal observables at $M_{Z'}=4$ TeV and the canonical ILC integrated luminosities 
250 fb${}^{-1}$, 500 fb${}^{-1}$, and 1000 fb${}^{-1}$ for 0.25, 0.5, and 1 TeV.}
\begin{center}
\begin{tabular}{cccc} 
\hline\hline
 & 250 GeV &  500 GeV & 1 TeV \\ 
\hline 
\multicolumn{4}{c}{$z\in[-1,1]$} \\
\hline
SSM&1.50&4.31&12.92\\
$\chi$&1.24&3.60&10.93\\
$\psi$&0.39&1.15&3.51\\
$\eta$&0.44&1.29&3.95\\
LRS&0.79&2.23&6.59\\
ALR&2.38&6.62&19.19\\
LH&0.50&1.40&4.13\\
SLH&0.17&0.51&1.56\\
$\g{U}(1)_{X,x=1}$&6.97&20.48&62.80\\
$\g{U}(1)_{X,x=-1.25}$&3.36&9.88&30.33\\
\hline
\multicolumn{4}{c}{$z\in[-0.9848,0.9848]$, ($10^\circ < \theta < 170^\circ$)} \\
\hline
SSM&1.48&4.23&12.69\\
$\chi$&1.23&3.56&10.81\\
$\psi$&0.39&1.13&3.45\\
$\eta$&0.43&1.27&3.90\\
LRS&0.77&2.19&6.48\\
ALR&2.35&6.54&18.98\\
LH&0.49&1.38&4.08\\
SLH&0.17&0.50&1.54\\
$\g{U}(1)_{X,x=1}$&6.88&20.25&62.10\\
$\g{U}(1)_{X,x=-1.25}$&3.32&9.78&30.01\\
\hline\hline
\end{tabular}
\end{center}
\end{table}

\section{Effects of $Z$--$Z'$ mixing}

The $Z$--$Z'$ mixing angle $\theta_0$ arises from the diagonalization of the mass matrix of neutral vector bosons. For a bunch of $Z'$ models, it was discussed in details in Ref. \cite{Langacker:1991pg}. In general, $\theta_0$ depends on the couplings and the vacuum expectation values of scalar fields responsible for the spontaneous breakdown of gauge symmetries. Because of different possibilities to introduce the scalar sector, additional parameters like the ratios of vacuum expectation values occur, and the mixing angle cannot be reduced to the couplings in Table \ref{Table:couplings} exclusively. So, it has to be considered as a free parameter in the $Z'$ models.

The $Z$--$Z'$ mixing angle $\theta_0$ can be written in the form 
\begin{equation}
\theta_0 ={\cal C}\frac{g_{Z'}}{g_Z}\frac{M^2_Z}{M^2_{Z'}}.
\label{mixingangle}
\end{equation}
where ${\cal C}\sim 1$ is a model dependent factor. Explicit factors ${\cal C}$ can be found in Ref. \cite{Langacker:1991pg}.

In presence of $Z$--$Z'$ mixing, the Lagrangian (\ref{lag}) becomes more complicated. Namely, the $Z$ and $Z'$ couplings must be substituted by
\begin{eqnarray}
&&v_f \to v_f \cos\theta_0 + v'_f\sin\theta_0,
\nonumber\\&&
a_f \to a_f \cos\theta_0 +a'_f\sin\theta_0,
\nonumber\\&&
v'_f \to v'_f \cos\theta_0 - v_f\sin\theta_0,
\nonumber\\&&
a'_f \to a'_f \cos\theta_0 -a_f\sin\theta_0.
\label{lagmixing}
\end{eqnarray}

Since both the parameters $\theta_0$ and $\mu$ behave like $\sim M_{Z'}^{-2}$, only the linear in $\theta_0$ term in the cross section could correct the results from the previous sections:
\begin{equation}
\frac{d\sigma}{dz}\simeq \frac{d\sigma^\mathrm{SM}}{dz} + \mu\, F_1(s,z,a',v') + \theta_0\, F_\mathrm{mix}(s,z,a',v').
\label{otdifmix}
\end{equation}
Let us notice that the current constraint on $\theta_0$ ($\le 10^{-4}$) means $\theta_0\le \mu$ for the ILC energies.
Using the same notations as in (\ref{F}) , (\ref{F2}), we can write $F_\mathrm{mix}$ in the following form:
\begin{eqnarray}
&&F_\mathrm{mix}(s,z,a',v')= -\frac{\alpha^{3/2}_\mathrm{em}g_{Z^\prime}\sqrt{\pi}}{32\sin^3\theta_W\cos^3\theta_W (s-M^2_Z)}
\nonumber\\&&
\quad\times
\Big\{z\,f_\mathrm{mix}\left[s,a'_e,2\epsilon(v'_e+a'_e\epsilon)\right] 
\nonumber\\&&\qquad\quad
+ \frac{1+z^2}{2}f_\mathrm{mix}\left[s,v'_e\epsilon,(a'_e+v'_e\epsilon)(1+\epsilon^2)\right]\Big\},
\label{Fmix}
\end{eqnarray}
with
\begin{equation}
f_\mathrm{mix}(s,x,y) = x (1 - \epsilon) (3 + \epsilon) +y  \frac{s}{s-M^2_Z}.
\label{fmix}
\end{equation}
For qualitative analysis, the small parameter $\epsilon$ can be omitted:
\begin{eqnarray}
&&F_\mathrm{mix}
\sim
\left[z\,f_\mathrm{mix}(s,a'_e,0) + \frac{1+z^2}{2}f_\mathrm{mix}(s,0,a'_e)\right].
\label{Fmix0}
\end{eqnarray}
As it is seen, the contribution from the $Z$--$Z'$ mixing is related to the axial-vector coupling. This fact is explained as follows. In the SM, the $Z$ vector coupling to charged leptons is suppressed (it is $\sim\epsilon$, see the SSM couplings with $s_W^2\simeq 1/4$). On the other hand, only the $Z$ exchange Feynman diagram contributes to $F_\mathrm{mix}$ (the $Z'$ exchange Feynman diagram $\sim\mu\,\epsilon$).

In order to estimate effects from $Z$--$Z'$ mixing let us compare (\ref{Fmix0}) with (\ref{F}). First of all, for $v'_e=0$ both the factors $F_1$ and $F_\mathrm{mix}$ contain the same angular dependency (up to $\epsilon^2$). This means that {\it in the models with suppressed vector coupling $v'_e$ the mixing angle corrects the estimated parameter $\mu$ and cannot be separated in the leading order in $M^{-2}_{Z'}$}. Thus, for the SSM, LRS, $\psi$, and SLH models, all the results from the previous sections remain unchanged after updating $\mu$:
\begin{equation}
\mu \to \mu^* = \mu + \theta_0 \frac{\sqrt{\pi\alpha}M^2_Z}{a'_e g_{Z'}\sin\theta_W\cos\theta_W (M^2_Z-s)}.
\label{mueff}
\end{equation}
The effective parameter $\mu^*$ is shown in Table \ref{Table:MuEff} for different ILC energies. In what follows, we will ignore these models discussing the $Z$--$Z'$ mixing.

\begin{table}
\caption{\label{Table:MuEff} The effective parameter $\mu^*$ in the models with suppressed $v'_e$.}
\begin{center}
\begin{tabular}{cccc} 
\hline\hline
 & 250 GeV &  500 GeV & 1 TeV \\ 
\hline 
SSM&$\mu-0.31\,\theta_0$ &$\mu-0.07\,\theta_0$ &$\mu-0.02\,\theta_0$\\
LRS&$\mu+0.42\,\theta_0$ &$\mu+0.09\,\theta_0$ &$\mu+0.02\,\theta_0$\\
$\psi$&$\mu+0.60\,\theta_0$ &$\mu+0.13\,\theta_0$ &$\mu+0.03\,\theta_0$\\
SLH&$\mu+0.90\,\theta_0$ &$\mu+0.20\,\theta_0$ &$\mu+0.05\,\theta_0$\\
\hline\hline
\end{tabular}
\end{center}
\end{table}

Second, {\it in the models with suppressed axial-vector coupling $a'_e\simeq 0$, the $Z$--$Z'$ mixing may be ignored.} Indeed, in accordance with (\ref{Fmix0}), $\theta_0 \simeq 0$. The most close to this case model from the considered list is the $\eta$ model.

The factors in the cross section has universal behavior with respect to the collision energy:
\[
F_\mathrm{mix} \sim s^{-1},\quad F_1 \sim \mathrm{const},\quad M_Z\ll\sqrt{s}<M_{Z'}.
\]
$F_\mathrm{mix}$ is less than 2.5\% of $F_1$ at $\sqrt{s}=1$ TeV. For $\sqrt{s}=500$ GeV, it exceeds 5\% of $F_1$ only for the LH model. These magnitudes are comparable with the systematic error for $F_1$, so $Z$-$Z'$ mixing does not influence the results obtained in the previous sections for $\sqrt{s}\ge 500$ GeV.

\begin{figure}
\includegraphics[width=0.45\textwidth]{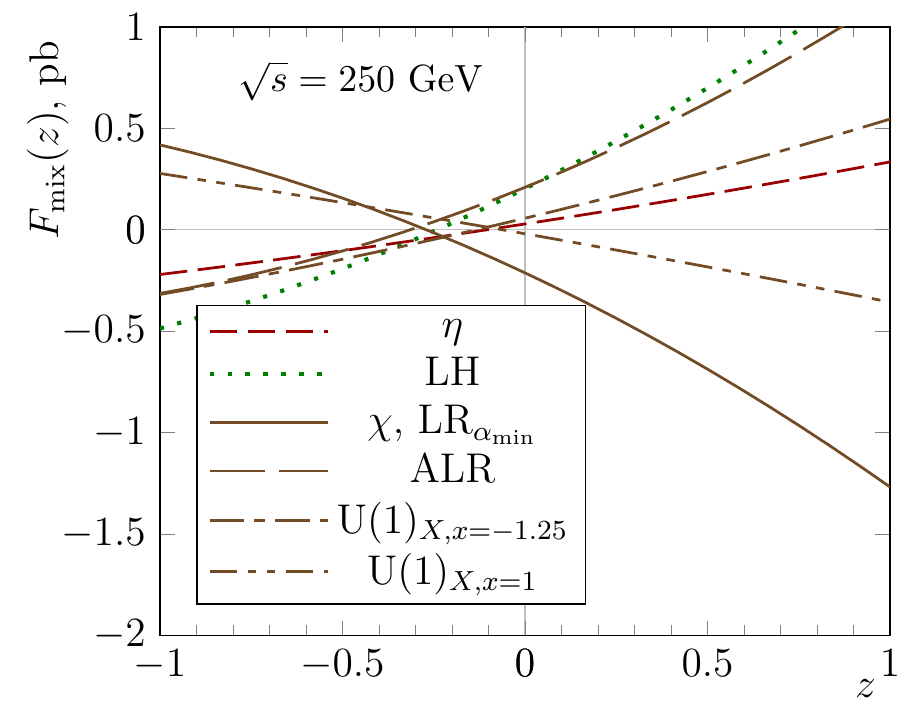}
\caption{The unpolarized factors $F_\mathrm{mix}(\sqrt{s},z)$ in pb at 250 GeV for the models with the mixing angle as a separate parameter. The factors are calculated up to 5\% systematic error. The plotted factors for the $\g{U}(1)_X$ model must be multiplied by 2.5.}\label{fig:trrk250mix}
\end{figure}

The angular dependence $F_\mathrm{mix}(z)$ is shown in Fig. \ref{fig:trrk250mix}. The odd-like shape is not surprising, since the effects of $Z$--$Z'$ mixing cannot be separated from $F_1$ in the models with odd-like $F_1$.

Visible effects of $Z$--$Z'$ occur at the lower ILC energy $\sqrt{s}=250$ GeV only. In this case there are optimal observables to select either $\mu$ or $\theta_0$.
The easiest way to derive them is either to adopt general formulas form \cite{Atwood:1991ka,Diehl:1993br} or to write the $\chi^2$ fit with several parameters. Let us denote the set of fitted parameters as $\gamma_i$, the corresponding factors in the differential cross section as $F_i$, and the weight function as $w_i$. In our case
\begin{eqnarray}
&&
\gamma_i = [\mu,\theta_0],
\nonumber\\&&
F_i = [F_1,F_\mathrm{mix}],
\nonumber\\&&
w_i = [w_\mu,w_{\theta_0}].
\label{mixparam}
\end{eqnarray}
The result is
\begin{eqnarray}
&&
\gamma_{i,\mathrm{ML}} = \int_\Omega dz\,w_{i,\mathrm{ML}}(z)\left(\frac{d\sigma}{dz}-\frac{d\sigma^\mathrm{SM}}{dz}\right),
\nonumber\\&&
w_{i,\mathrm{ML}} = \sum\limits_{j} \tilde{c}^{-1}_{ij}\,\frac{F_j(z)}{d\sigma^\mathrm{SM}/dz},
\label{mixw}
\end{eqnarray}
where $\tilde{c}^{-1}_{ij}$ is the inverse matrix for
\begin{equation}
\tilde{c}_{ij} = \int_\Omega dz\ \frac{F_i(z)F_j(z)}{d\sigma^\mathrm{SM}/dz}.
\label{cij}
\end{equation}
The numerical values of $\tilde{c}^{-1}_{ij}$ are given in Table \ref{Table:cinv}.
It is easy to see that (\ref{mixw}) gives (\ref{MLobserv}) for the single parameter $\mu$.

\begin{table}
\caption{\label{Table:cinv} The numeric values of matrix $\tilde{c}^{-1}_{ij}$. The complete phase space is suggested ($z\in[-1,1]$).}
\begin{center}
\begin{tabular}{cccc} 
\hline\hline
 & $\tilde{c}^{-1}_{11}$ &  $\tilde{c}^{-1}_{12}$ & $\tilde{c}^{-1}_{22}$ \\ 
\hline 
\multicolumn{4}{c}{250 GeV} \\
\hline
$\chi$ & 0.0576617 & -0.176493 & 2.32241\\
$\eta$ & 0.355642 & -0.0730611 & 13.4809\\
ALR & 0.0152012 & 0.097004 & 2.97555\\
LH & 0.631465 & 1.142 & 3.64662\\
$\g{U}(1)_{X,x=1}$& 0.00143955 & 0.0075562 & 1.53929\\
$\g{U}(1)_{X,x=-1.25}$& 0.00609376 & 0.00842994 & 0.956269\\
\hline 
\multicolumn{4}{c}{500 GeV} \\
\hline
$\chi$ & 0.0130213 & -0.157882 & 11.0761\\
$\eta$ & 0.0841745 & 0.0543964 & 63.7052\\
ALR & 0.0036781 & 0.0836812 & 14.2823\\
LH & 0.149855 & 1.1523 & 16.9333\\
$\g{U}(1)_{X,x=1}$ & 0.000334045 & 0.00370235 & 7.17784\\
$\g{U}(1)_{X,x=-1.25}$ & 0.00145713 & 0.0117845 & 4.56711\\
\hline 
\multicolumn{4}{c}{1 TeV} \\
\hline
$\chi$ & 0.00296117 & -0.131345 & 44.4095\\
$\eta$ & 0.0198895 & 0.206536 & 254.214\\
ALR & 0.000909918 & 0.0655109 & 56.9887\\
LH & 0.0355517 & 1.08269 & 67.0021\\
$\g{U}(1)_{X,x=1}$ & 0.0000779599 & -0.000906613 & 28.3105\\
$\g{U}(1)_{X,x=-1.25}$ & 0.000346823 & 0.0152999 & 18.4215\\
\hline\hline
\end{tabular}
\end{center}
\end{table}

In terms of the Hilbert space from Sec. \ref{secOO},
\begin{eqnarray}
&&
\tilde{c}_{ij} = \left\langle \frac{F_i}{d\sigma^\mathrm{SM}/dz},\ \frac{F_j}{d\sigma^\mathrm{SM}/dz} \right\rangle,
\nonumber\\&&
\left\langle w_{i,\mathrm{ML}},\ \frac{F_j}{d\sigma^\mathrm{SM}/dz} \right\rangle = \left\{\begin{array}{cl}1,&i=j,\\0,&i\not=j.\end{array}\right.
\label{mixwHilbert}
\end{eqnarray}
The last equation corresponds to (\ref{MLnorm}). First of all, it means that {\it the weight function for a given parameter must be orthogonal to all the rest $Z'$ factors in the differential cross section},
\begin{equation}
\int_\Omega dz\ w_{i,\mathrm{ML}}(z) F_{j\not=i}(z) = 0.
\label{orth}
\end{equation}
With this projection of $F_i$ to the axis orthogonal to any other factor, we obtain the result exactly corresponding to the one-dimensional optimization from Sec. \ref{secOO}. This is the definition of the optimal observable through {\it conditional maximization of the signal-to-uncertainty ratio} (\ref{otnoso}) with the additional constraints to exclude contributions from other unknown parameters \cite{Gulov:2013kpa}. 

\begin{figure}
\includegraphics[width=0.45\textwidth]{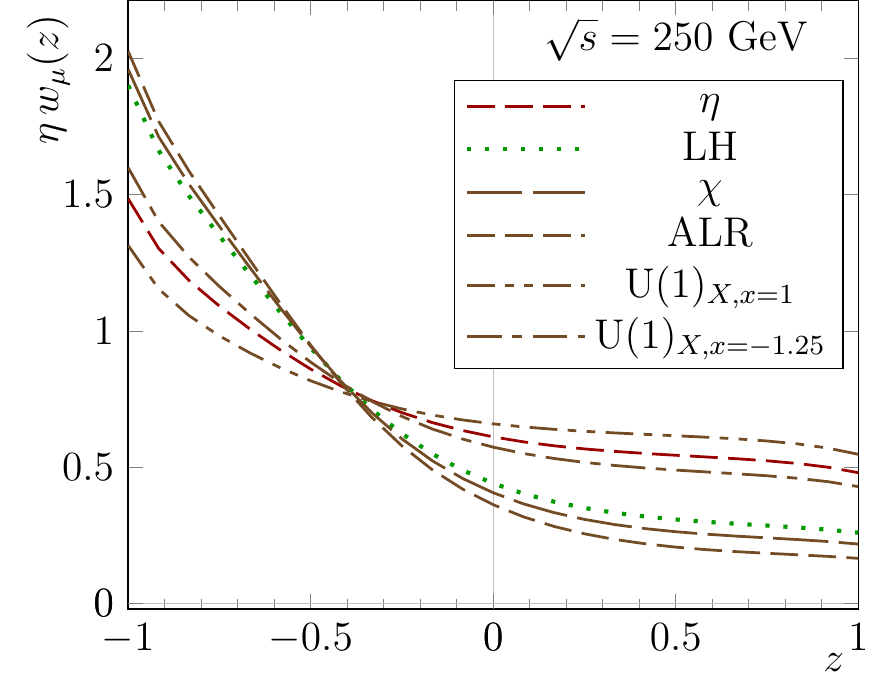}
\includegraphics[width=0.45\textwidth]{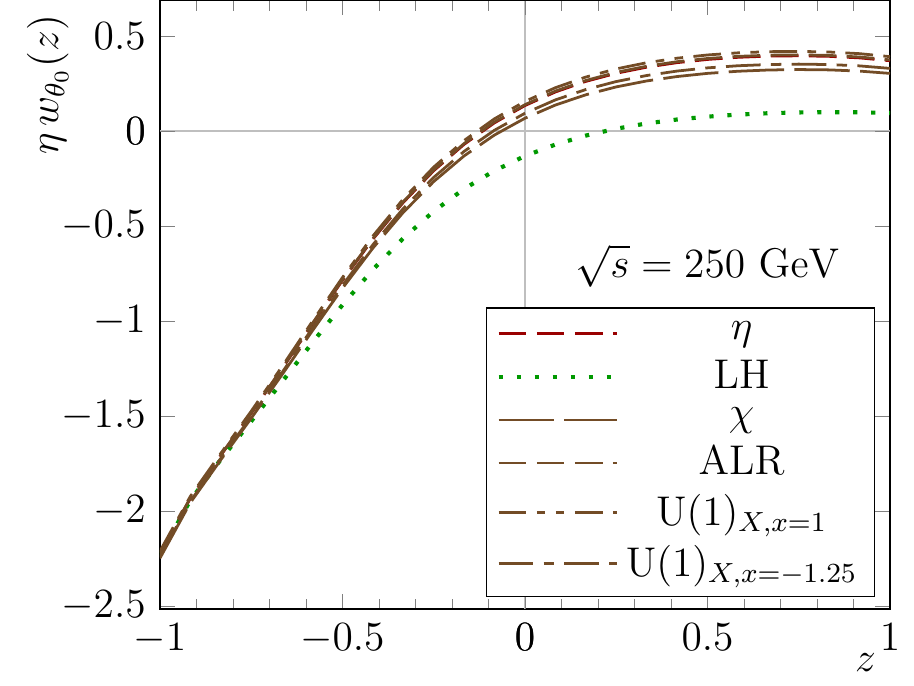}
\caption{The weight functions $\eta\,w(\sqrt{s},z)$ to measure $M_{Z'}$ and $\theta_0$ at the ILC energies. The universal normalization $\int_{-1}^1 w^2 dz=1$ is used to compare different models. The sign $\eta=-1$ for the $\chi$ and $\g{U}(1)_{X,x=1}$ models , and $\eta=1$ otherwise. The lines correspond to the SM from ZFITTER. The complete phase space $z\in[-1,1]$ is used.}
\label{fig:wmix}
\end{figure}

The weight functions to measure separately $\mu$ and $\theta_0$ at $\sqrt{s}=250$ GeV are shown in Fig. \ref{fig:wmix}. If the weight function for $\mu$ is close to the total cross section in absence of the $Z$--$Z'$ mixing, we see no qualitative changes of the shape of $w_\mu$ with $\theta_0$ taken into account. This can be explained by the fact that such models have initially approximately orthogonal factors $F_1$ (even) and $F_\mathrm{mix}$ (odd). However, the common tendency is the growing weight of the backward bins.

The weight function to measure the $Z$--$Z'$ mixing angle has always an odd-like shape. It is close to the weight function for $\mu^*$ in the models with suppressed vector coupling of $Z'$ boson to charged leptons.

For the models with the $Z'$ couplings to left-handed chiral states of charged leptons only (the LH model), the situation becomes more interesting. In this case we have three completely different optimal observables: two observables to measure $\mu$ with either absence or presence of the $Z$--$Z'$ mixing and one observable to measure the mixing angle. Their application to data would allow to obtain a lot of information about the $Z$--$Z'$ mixing angle.

\begin{figure}
\includegraphics[width=0.45\textwidth]{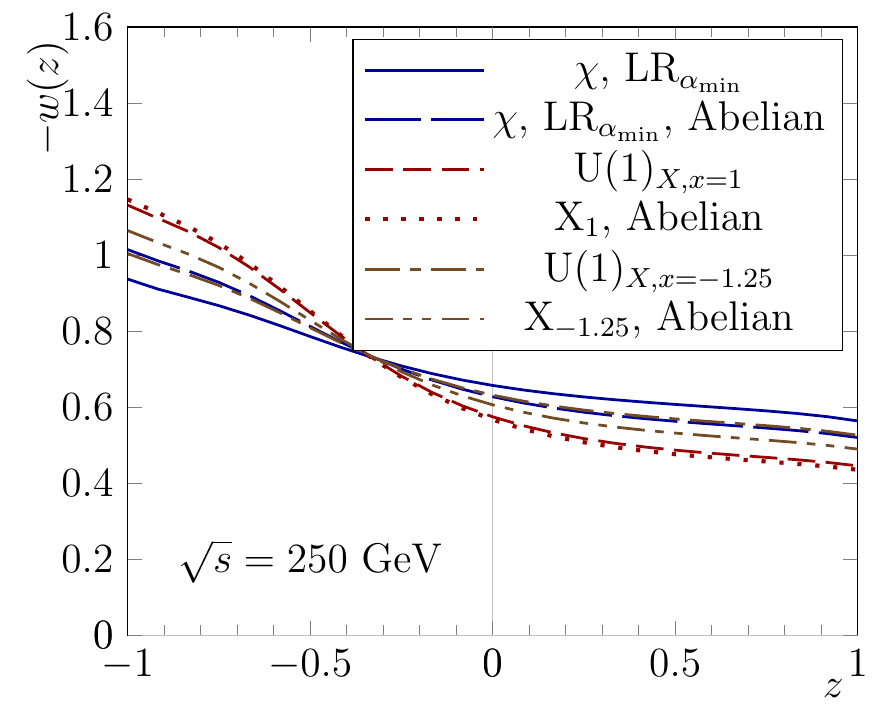}
\caption{The weight functions $-w(\sqrt{s},z)$ to measure $\mu$ in the Abelian models with $\theta_0$ from (\ref{mixAb}). The universal normalization $\int_{-1}^1 w^2 dz=1$ is used to compare different models. The shorthand notations $\mathrm{X}_{-1.25}$ and $\mathrm{X}_{1}$ are used for $\g{U}(1)_{X,x=-1.25}$ and $\g{U}(1)_{X,x=-1.25}$ models.}
\label{fig:wmixAbelian}
\end{figure}

Another possibility to account for the $Z$--$Z'$ mixing is some model independent definitions of $\theta_0$. For instance, let us consider the Abelian $Z'$ boson \cite{Gulov:2000eh} including, in particular, the $\chi$, LR, and $\g{U}(1)_{X}$ models. In this case there is a relation inspired by the renormalization group equations below the threshold of $Z'$ decoupling:
\begin{equation}
\theta_0 \simeq \frac{4 g_{Z'} a'_e \sin\theta_W\cos\theta_W}{\sqrt{4 \pi \alpha_\mathrm{em}}} \frac{M^2_Z}{M^2_{Z'}}.
\label{mixAb}
\end{equation}
The optimal observables to measure Abelian $Z'$ couplings were discussed in \cite{Gulov:2013kpa}. It was also shown in Ref. \cite{Pevzner:2018chl} for the Abelian $Z'$ boson that the mixing angle plays an important role in calculations at the $Z'$ peak. Since $\theta_0$ is expressed in terms of the $Z'$ couplings, no additional parameters occur. However, factor $F_1$ has to be corrected to include contribution from the $Z$--$Z'$ mixing:
\begin{equation}
F_1 \to F_1 - \frac{4 g_{Z'} a'_e \sin\theta_W\cos\theta_W}{\sqrt{4 \pi \alpha_\mathrm{em}}} F_\mathrm{mix}.
\label{factorAb}
\end{equation}
The corresponding weight functions are compared with the case of absent mixing in Fig. \ref{fig:wmixAbelian}. As it is seen, the mixing angle affects the weight function up to 10\% at $\sqrt{s}=$ 250 GeV. For higher collision energies, the contribution from $\theta_0$ decreases essentially.

Finally, let us remind that there is no chance to separate and measure the $Z$--$Z'$ mixing angle in the models with suppressed vector $Z'$ couplings to charged leptons (SSM, $\psi$, LRS, SLH) as well as for collision energies $\sqrt{s}\ge500$ GeV within ILC experiments. Thus, the role of mixing is important at energies $\sim 250$ GeV.

\section{Discussion}

In the paper we have considered a set of popular $Z'$ models in the annihilation leptonic process. We have investigated weighted integrated cross sections with the best ratio of $Z'$ signal to statistical uncertainty (the optimal observables). They define uniquely the weight of every event in the phase space of the final particles. Then, all the available events can be summed up with the weights without intermediate aggregation into bins with respect to the scattering angle. The CL of the signal corresponds to the highest possible level allowed by the integrated luminosity. 

We have shown that the optimal observables are equivalent to the $\chi^2$ fit of the differential cross section. In this regard, it is a unique integration scheme of the cross section which leads to no losses of information encoded in the differential cross section. The optimal observables can be implemented as a simple and convenient alternative to the complete analysis of the differential cross section: it could simply accumulate events from the start of an experiment without additional manipulation of data leading directly to the ML estimator for the $Z'$ mass.

It is interesting to compare the estimates based on the optimal observables with the popular approach taking into consideration the forward-backward asymmetry $A_{FB}$. Such an analysis can explain a lot of details of application of $A_{FB}$ within the $Z'$ models. In Ref. \cite{Han:2013mra} the 95\% exclusion reaches for the $Z'$ mass were calculated by means of $A_{FB}$ assuming polarizations $P^-=0.8,~P^+=0.3$ at 0.5 TeV and $P^-=0.8,~P^+=0.2$ at 1 TeV. We choose the same settings and calculate exclusion reaches based on the optimal observables. First of all, substituting the deviation from the SM in (\ref{MLobserv}) by the value predicted by the $Z'$ model,
\begin{equation}
\frac{d\sigma}{dz}-\frac{d\sigma^\mathrm{SM}}{dz}\simeq  \frac{M^2_Z}{s-M^2_{Z'}}\ F_1(z),
\end{equation}
we obtain the obvious ML estimate of the model parameter (\ref{mu}). Then, specifying the 95\% confidence level, $N_\mathrm{CL}=2$, and solving (\ref{NsignalC}) for the $Z'$ mass, we obtain {\it the exclusion reach},
\begin{equation}
M^2_{Z',95\%}\simeq s + \frac{M^2_Z}{2}\sqrt{{\cal L}\int_\Omega dz\ \frac{F_1^2(z)}{d\sigma^\mathrm{SM}/dz}},
\label{ereach}
\end{equation}
since any lower value of the $Z'$ mass will be detected as the signal at the considered CL.

\begin{table}
\caption{\label{negmix} The exclusion reach on $M_{Z'}$ (TeV) at 95\% CL from the optimal observable. The correspondent exclusion reach from the forward-backward asymmetry from Ref. \cite{Han:2013mra} are also presented. The polarizations $P^-=0.8,~P^+=-0.3$ at 500 GeV and $P^-=0.8,~P^+=-0.2$ at 1 TeV are assumed. The results in \cite{Han:2013mra} are rounded up to 0.5 TeV.}
\begin{center}
\begin{tabular}{cccccc} 
\hline\hline
$\sqrt{s}$ & 250 GeV & \multicolumn{2}{c}{500 GeV} & \multicolumn{2}{c}{1 TeV} \\ 
& Eq.(\ref{ereach})   & Ref.\cite{Han:2013mra} & Eq.(\ref{ereach}) & Ref.\cite{Han:2013mra} &  Eq.(\ref{ereach}) \\ 
\hline 
SSM   & 3.4 & 5.5 & 5.8 & 9.8 & 9.8 \\
$\psi$ & 1.8 & 2.7 & 3.0 & 4.7 & 5.2 \\
LRS    & 2.5 & 3.7 & 4.2 & 6.5 & 7.0 \\
SLH    & 1.2 & --&	2.0 & --& 3.5 \\
$\chi$ & 3.1 & 1.6 & 5.3 & 3 & 9.1 \\
$\eta$ & 1.9 & 1.7 & 3.2 & 3 & 5.5 \\
ALR    & 4.3 & --& 7.2 &--& 12.0 \\
LH      & 2.0 &--& 3.3 & --& 5.6 \\ 
$\g{U}(1)_{X,x=1}$&7.4&--&12.6&--&21.6\\
$\g{U}(1)_{X,x=-1.25}$&5.2&--&8.8&--&15.0\\
\hline\hline
\end{tabular}
\end{center}
\end{table} 

In Table \ref{negmix} we compare the exclusion reaches from the optimal observables and from the forward-backward asymmetries \cite{Han:2013mra}. We take the same interval of scattering angles $10^\circ < \theta < 170^\circ$, which corresponds to $z\in[-0.9848,0.9848]$. The exclusion reaches from the optimal observables are always higher than in any other scheme, since this approach is more efficient from the statistical point of view. However, the difference is not uniform over the $Z'$ models. In case of the SSM, $\psi$, and LRS model, the weight function is closer to the forward-backward integration scheme, so the exclusion reach increases up to 10\%. Actually, the forward-backward asymmetry might be good to search for these models. On the other hand, the weight functions of $\chi$ and $\eta$ models are far away from odd shapes. The exclusion reaches from $A_{FB}$ are weak in this case, and the forward-backward asymmetry seems to be insufficient to fit data. Indeed, the optimal observables increase the exclusion reaches up to 200\%. 

As a concluding remark, we notice that the optimal observable shows the best fit of data. It allows to obtain the ML estimator of the $Z'$ mass taking into account the distribution of events over the phase space of the final particles with no loss of information. The approach can be applied to treat experimental data even for small samples when the aggregation of events into detailed differential cross section is practically impossible. Such a feature would be very useful at the start of experiment. Indeed, (\ref{MLobserv}) can be rewritten as
\begin{equation}
\frac{1}{s-M^{2}_{Z',\mathrm{ML}}} \simeq \sum\limits_{i\in\mathrm{events}}\frac{w_{\mathrm{ML},i}}{M^2_Z \cal L}  - \int\limits_{-1}^{1}dz\,\frac{w_\mathrm{ML}(z)}{M^2_Z}\,\frac{d\sigma^\mathrm{SM}}{dz},
\label{MLobservEvents}
\end{equation}
where the sum runs over all the observed events, $w_{\mathrm{ML},i}$ corresponds to the measured scattering angle, ${\cal L}$ means the actual luminosity taking into account the event acceptance rate, and the integral is the SM background calculated, for instance, by the event generator for the actual detector. Thus, the optimal observable can be considered as a technique of continuous accumulation of events with on the fly fit of the $Z'$ mass from the implied differential cross section without actual construction of the differential cross section. No signal from the optimal observable would mean that there is no sense in further searching for the $Z'$ model in the collected data.

Introducing the acceptance rates in the phase space and other minor technical improvements are left beyond the scope of this paper. They could be revisited at the stage of practical implementation of the optimal observables at future lepton colliders.

\bibliography{bibliography}

\begin{thebibliography}{34}%
\makeatletter
\providecommand \@ifxundefined [1]{%
 \@ifx{#1\undefined}
}%
\providecommand \@ifnum [1]{%
 \ifnum #1\expandafter \@firstoftwo
 \else \expandafter \@secondoftwo
 \fi
}%
\providecommand \@ifx [1]{%
 \ifx #1\expandafter \@firstoftwo
 \else \expandafter \@secondoftwo
 \fi
}%
\providecommand \natexlab [1]{#1}%
\providecommand \enquote  [1]{``#1''}%
\providecommand \bibnamefont  [1]{#1}%
\providecommand \bibfnamefont [1]{#1}%
\providecommand \citenamefont [1]{#1}%
\providecommand \href@noop [0]{\@secondoftwo}%
\providecommand \href [0]{\begingroup \@sanitize@url \@href}%
\providecommand \@href[1]{\@@startlink{#1}\@@href}%
\providecommand \@@href[1]{\endgroup#1\@@endlink}%
\providecommand \@sanitize@url [0]{\catcode `\\12\catcode `\$12\catcode
  `\&12\catcode `\#12\catcode `\^12\catcode `\_12\catcode `\%12\relax}%
\providecommand \@@startlink[1]{}%
\providecommand \@@endlink[0]{}%
\providecommand \url  [0]{\begingroup\@sanitize@url \@url }%
\providecommand \@url [1]{\endgroup\@href {#1}{\urlprefix }}%
\providecommand \urlprefix  [0]{URL }%
\providecommand \Eprint [0]{\href }%
\providecommand \doibase [0]{http://dx.doi.org/}%
\providecommand \selectlanguage [0]{\@gobble}%
\providecommand \bibinfo  [0]{\@secondoftwo}%
\providecommand \bibfield  [0]{\@secondoftwo}%
\providecommand \translation [1]{[#1]}%
\providecommand \BibitemOpen [0]{}%
\providecommand \bibitemStop [0]{}%
\providecommand \bibitemNoStop [0]{.\EOS\space}%
\providecommand \EOS [0]{\spacefactor3000\relax}%
\providecommand \BibitemShut  [1]{\csname bibitem#1\endcsname}%
\let\auto@bib@innerbib\@empty
\bibitem [{\citenamefont {Baer}\ \emph {et~al.}(2013)\citenamefont {Baer},
  \citenamefont {Barklow}, \citenamefont {Fujii}, \citenamefont {Gao},
  \citenamefont {Hoang}, \citenamefont {Kanemura}, \citenamefont {List},
  \citenamefont {Logan}, \citenamefont {Nomerotski}, \citenamefont {Perelstein}
  \emph {et~al.}}]{Baer:2013cma}%
  \BibitemOpen
  \bibfield  {author} {\bibinfo {author} {\bibfnamefont {H.}~\bibnamefont
  {Baer}}, \bibinfo {author} {\bibfnamefont {T.}~\bibnamefont {Barklow}},
  \bibinfo {author} {\bibfnamefont {K.}~\bibnamefont {Fujii}}, \bibinfo
  {author} {\bibfnamefont {Y.}~\bibnamefont {Gao}}, \bibinfo {author}
  {\bibfnamefont {A.}~\bibnamefont {Hoang}}, \bibinfo {author} {\bibfnamefont
  {S.}~\bibnamefont {Kanemura}}, \bibinfo {author} {\bibfnamefont
  {J.}~\bibnamefont {List}}, \bibinfo {author} {\bibfnamefont {H.~E.}\
  \bibnamefont {Logan}}, \bibinfo {author} {\bibfnamefont {A.}~\bibnamefont
  {Nomerotski}}, \bibinfo {author} {\bibfnamefont {M.}~\bibnamefont
  {Perelstein}},  \emph {et~al.},\ }\href@noop {} {\  (\bibinfo {year}
  {2013})},\ \Eprint {http://arxiv.org/abs/1306.6352} {arXiv:1306.6352
  [hep-ph]} \BibitemShut {NoStop}%
\bibitem [{\citenamefont {Weiglein}\ \emph {et~al.}(2006)\citenamefont
  {Weiglein} \emph {et~al.}}]{Weiglein:2004hn}%
  \BibitemOpen
  \bibfield  {author} {\bibinfo {author} {\bibfnamefont {G.}~\bibnamefont
  {Weiglein}} \emph {et~al.} (\bibinfo {collaboration} {LHC/LC Study Group}),\
  }\href {\doibase 10.1016/j.physrep.2005.12.003} {\bibfield  {journal}
  {\bibinfo  {journal} {Phys. Rept.}\ }\textbf {\bibinfo {volume} {426}},\
  \bibinfo {pages} {47} (\bibinfo {year} {2006})},\ \Eprint
  {http://arxiv.org/abs/hep-ph/0410364} {arXiv:hep-ph/0410364 [hep-ph]}
  \BibitemShut {NoStop}%
\bibitem [{\citenamefont {Aaboud}\ \emph {et~al.}(2017)\citenamefont {Aaboud}
  \emph {et~al.}}]{Aaboud:2017buh}%
  \BibitemOpen
  \bibfield  {author} {\bibinfo {author} {\bibfnamefont {M.}~\bibnamefont
  {Aaboud}} \emph {et~al.} (\bibinfo {collaboration} {ATLAS}),\ }\href
  {\doibase 10.1007/JHEP10(2017)182} {\bibfield  {journal} {\bibinfo  {journal}
  {JHEP}\ }\textbf {\bibinfo {volume} {10}},\ \bibinfo {pages} {182} (\bibinfo
  {year} {2017})},\ \Eprint {http://arxiv.org/abs/1707.02424} {arXiv:1707.02424
  [hep-ex]} \BibitemShut {NoStop}%
\bibitem [{\citenamefont {Sirunyan}\ \emph {et~al.}(2018)\citenamefont
  {Sirunyan} \emph {et~al.}}]{Sirunyan:2018exx}%
  \BibitemOpen
  \bibfield  {author} {\bibinfo {author} {\bibfnamefont {A.~M.}\ \bibnamefont
  {Sirunyan}} \emph {et~al.} (\bibinfo {collaboration} {CMS}),\ }\href
  {\doibase 10.1007/JHEP06(2018)120} {\bibfield  {journal} {\bibinfo  {journal}
  {JHEP}\ }\textbf {\bibinfo {volume} {06}},\ \bibinfo {pages} {120} (\bibinfo
  {year} {2018})},\ \Eprint {http://arxiv.org/abs/1803.06292} {arXiv:1803.06292
  [hep-ex]} \BibitemShut {NoStop}%
\bibitem [{\citenamefont {Gulov}(2014)}]{Gulov:2013kpa}%
  \BibitemOpen
  \bibfield  {author} {\bibinfo {author} {\bibfnamefont {A.}~\bibnamefont
  {Gulov}},\ }\href {\doibase 10.1142/S0217751X14501619} {\bibfield  {journal}
  {\bibinfo  {journal} {Int. J. Mod. Phys.}\ }\textbf {\bibinfo {volume}
  {A29}},\ \bibinfo {pages} {1450161} (\bibinfo {year} {2014})},\ \Eprint
  {http://arxiv.org/abs/1308.4837} {arXiv:1308.4837 [hep-ph]} \BibitemShut
  {NoStop}%
\bibitem [{\citenamefont {Atwood}\ and\ \citenamefont
  {Soni}(1992)}]{Atwood:1991ka}%
  \BibitemOpen
  \bibfield  {author} {\bibinfo {author} {\bibfnamefont {D.}~\bibnamefont
  {Atwood}}\ and\ \bibinfo {author} {\bibfnamefont {A.}~\bibnamefont {Soni}},\
  }\href {\doibase 10.1103/PhysRevD.45.2405} {\bibfield  {journal} {\bibinfo
  {journal} {Phys. Rev.}\ }\textbf {\bibinfo {volume} {D45}},\ \bibinfo {pages}
  {2405} (\bibinfo {year} {1992})}\BibitemShut {NoStop}%
\bibitem [{\citenamefont {Diehl}\ and\ \citenamefont
  {Nachtmann}(1994)}]{Diehl:1993br}%
  \BibitemOpen
  \bibfield  {author} {\bibinfo {author} {\bibfnamefont {M.}~\bibnamefont
  {Diehl}}\ and\ \bibinfo {author} {\bibfnamefont {O.}~\bibnamefont
  {Nachtmann}},\ }\href {\doibase 10.1007/BF01555899} {\bibfield  {journal}
  {\bibinfo  {journal} {Z. Phys.}\ }\textbf {\bibinfo {volume} {C62}},\
  \bibinfo {pages} {397} (\bibinfo {year} {1994})}\BibitemShut {NoStop}%
\bibitem [{\citenamefont {Diehl}\ \emph {et~al.}(2003)\citenamefont {Diehl},
  \citenamefont {Nachtmann},\ and\ \citenamefont {Nagel}}]{Diehl:2002nj}%
  \BibitemOpen
  \bibfield  {author} {\bibinfo {author} {\bibfnamefont {M.}~\bibnamefont
  {Diehl}}, \bibinfo {author} {\bibfnamefont {O.}~\bibnamefont {Nachtmann}}, \
  and\ \bibinfo {author} {\bibfnamefont {F.}~\bibnamefont {Nagel}},\ }\href
  {\doibase 10.1140/epjc/s2002-01096-y} {\bibfield  {journal} {\bibinfo
  {journal} {Eur. Phys. J.}\ }\textbf {\bibinfo {volume} {C27}},\ \bibinfo
  {pages} {375} (\bibinfo {year} {2003})},\ \Eprint
  {http://arxiv.org/abs/hep-ph/0209229} {arXiv:hep-ph/0209229 [hep-ph]}
  \BibitemShut {NoStop}%
\bibitem [{\citenamefont {Aad}\ \emph {et~al.}(2016)\citenamefont {Aad} \emph
  {et~al.}}]{Aad:2016nal}%
  \BibitemOpen
  \bibfield  {author} {\bibinfo {author} {\bibfnamefont {G.}~\bibnamefont
  {Aad}} \emph {et~al.} (\bibinfo {collaboration} {ATLAS}),\ }\href {\doibase
  10.1140/epjc/s10052-016-4499-5} {\bibfield  {journal} {\bibinfo  {journal}
  {Eur. Phys. J.}\ }\textbf {\bibinfo {volume} {C76}},\ \bibinfo {pages} {658}
  (\bibinfo {year} {2016})},\ \Eprint {http://arxiv.org/abs/1602.04516}
  {arXiv:1602.04516 [hep-ex]} \BibitemShut {NoStop}%
\bibitem [{\citenamefont {Hewett}\ and\ \citenamefont
  {Rizzo}(1989)}]{Hewett:1988xc}%
  \BibitemOpen
  \bibfield  {author} {\bibinfo {author} {\bibfnamefont {J.~L.}\ \bibnamefont
  {Hewett}}\ and\ \bibinfo {author} {\bibfnamefont {T.~G.}\ \bibnamefont
  {Rizzo}},\ }\href {\doibase 10.1016/0370-1573(89)90071-9} {\bibfield
  {journal} {\bibinfo  {journal} {Phys. Rept.}\ }\textbf {\bibinfo {volume}
  {183}},\ \bibinfo {pages} {193} (\bibinfo {year} {1989})}\BibitemShut
  {NoStop}%
\bibitem [{\citenamefont {Leike}(1999)}]{Leike:1998wr}%
  \BibitemOpen
  \bibfield  {author} {\bibinfo {author} {\bibfnamefont {A.}~\bibnamefont
  {Leike}},\ }\href {\doibase 10.1016/S0370-1573(98)00133-1} {\bibfield
  {journal} {\bibinfo  {journal} {Phys. Rept.}\ }\textbf {\bibinfo {volume}
  {317}},\ \bibinfo {pages} {143} (\bibinfo {year} {1999})},\ \Eprint
  {http://arxiv.org/abs/hep-ph/9805494} {arXiv:hep-ph/9805494 [hep-ph]}
  \BibitemShut {NoStop}%
\bibitem [{\citenamefont {Rizzo}(2006)}]{Rizzo:2006nw}%
  \BibitemOpen
  \bibfield  {author} {\bibinfo {author} {\bibfnamefont {T.~G.}\ \bibnamefont
  {Rizzo}},\ }in\ \href
  {http://www-public.slac.stanford.edu/sciDoc/docMeta.aspx?slacPubNumber=slac-pub-12129}
  {\emph {\bibinfo {booktitle} {{Proceedings of Theoretical Advanced Study
  Institute in Elementary Particle Physics : Exploring New Frontiers Using
  Colliders and Neutrinos (TASI 2006): Boulder, Colorado, June 4-30, 2006}}}}\
  (\bibinfo {year} {2006})\ pp.\ \bibinfo {pages} {537--575},\ \Eprint
  {http://arxiv.org/abs/hep-ph/0610104} {arXiv:hep-ph/0610104 [hep-ph]}
  \BibitemShut {NoStop}%
\bibitem [{\citenamefont {Langacker}(2009)}]{Langacker:2008yv}%
  \BibitemOpen
  \bibfield  {author} {\bibinfo {author} {\bibfnamefont {P.}~\bibnamefont
  {Langacker}},\ }\href {\doibase 10.1103/RevModPhys.81.1199} {\bibfield
  {journal} {\bibinfo  {journal} {Rev. Mod. Phys.}\ }\textbf {\bibinfo {volume}
  {81}},\ \bibinfo {pages} {1199} (\bibinfo {year} {2009})},\ \Eprint
  {http://arxiv.org/abs/0801.1345} {arXiv:0801.1345 [hep-ph]} \BibitemShut
  {NoStop}%
\bibitem [{\citenamefont {Gursey}\ \emph {et~al.}(1976)\citenamefont {Gursey},
  \citenamefont {Ramond},\ and\ \citenamefont {Sikivie}}]{Gursey:1975ki}%
  \BibitemOpen
  \bibfield  {author} {\bibinfo {author} {\bibfnamefont {F.}~\bibnamefont
  {Gursey}}, \bibinfo {author} {\bibfnamefont {P.}~\bibnamefont {Ramond}}, \
  and\ \bibinfo {author} {\bibfnamefont {P.}~\bibnamefont {Sikivie}},\ }\href
  {\doibase 10.1016/0370-2693(76)90417-2} {\bibfield  {journal} {\bibinfo
  {journal} {Phys. Lett.}\ }\textbf {\bibinfo {volume} {60B}},\ \bibinfo
  {pages} {177} (\bibinfo {year} {1976})}\BibitemShut {NoStop}%
\bibitem [{\citenamefont {Pati}\ and\ \citenamefont
  {Salam}(1974)}]{Pati:1974yy}%
  \BibitemOpen
  \bibfield  {author} {\bibinfo {author} {\bibfnamefont {J.~C.}\ \bibnamefont
  {Pati}}\ and\ \bibinfo {author} {\bibfnamefont {A.}~\bibnamefont {Salam}},\
  }\href {\doibase 10.1103/PhysRevD.10.275, 10.1103/PhysRevD.11.703.2}
  {\bibfield  {journal} {\bibinfo  {journal} {Phys. Rev.}\ }\textbf {\bibinfo
  {volume} {D10}},\ \bibinfo {pages} {275} (\bibinfo {year} {1974})},\ \bibinfo
  {note} {[Erratum: Phys. Rev.D11,703(1975)]}\BibitemShut {NoStop}%
\bibitem [{\citenamefont {Ma}(1987)}]{Ma:1986we}%
  \BibitemOpen
  \bibfield  {author} {\bibinfo {author} {\bibfnamefont {E.}~\bibnamefont
  {Ma}},\ }\href {\doibase 10.1103/PhysRevD.36.274} {\bibfield  {journal}
  {\bibinfo  {journal} {Phys. Rev.}\ }\textbf {\bibinfo {volume} {D36}},\
  \bibinfo {pages} {274} (\bibinfo {year} {1987})}\BibitemShut {NoStop}%
\bibitem [{\citenamefont {Ashry}\ and\ \citenamefont
  {Khalil}(2015)}]{Ashry:2013loa}%
  \BibitemOpen
  \bibfield  {author} {\bibinfo {author} {\bibfnamefont {M.}~\bibnamefont
  {Ashry}}\ and\ \bibinfo {author} {\bibfnamefont {S.}~\bibnamefont {Khalil}},\
  }\href {\doibase 10.1103/PhysRevD.96.059901, 10.1103/PhysRevD.91.015009}
  {\bibfield  {journal} {\bibinfo  {journal} {Phys. Rev.}\ }\textbf {\bibinfo
  {volume} {D91}},\ \bibinfo {pages} {015009} (\bibinfo {year} {2015})},\
  \bibinfo {note} {[Addendum: Phys. Rev.D96,no.5,059901(2017)]},\ \Eprint
  {http://arxiv.org/abs/1310.3315} {arXiv:1310.3315 [hep-ph]} \BibitemShut
  {NoStop}%
\bibitem [{\citenamefont {Arkani-Hamed}\ \emph {et~al.}(2002)\citenamefont
  {Arkani-Hamed}, \citenamefont {Cohen}, \citenamefont {Katz},\ and\
  \citenamefont {Nelson}}]{ArkaniHamed:2002qy}%
  \BibitemOpen
  \bibfield  {author} {\bibinfo {author} {\bibfnamefont {N.}~\bibnamefont
  {Arkani-Hamed}}, \bibinfo {author} {\bibfnamefont {A.~G.}\ \bibnamefont
  {Cohen}}, \bibinfo {author} {\bibfnamefont {E.}~\bibnamefont {Katz}}, \ and\
  \bibinfo {author} {\bibfnamefont {A.~E.}\ \bibnamefont {Nelson}},\ }\href
  {\doibase 10.1088/1126-6708/2002/07/034} {\bibfield  {journal} {\bibinfo
  {journal} {JHEP}\ }\textbf {\bibinfo {volume} {07}},\ \bibinfo {pages} {034}
  (\bibinfo {year} {2002})},\ \Eprint {http://arxiv.org/abs/hep-ph/0206021}
  {arXiv:hep-ph/0206021 [hep-ph]} \BibitemShut {NoStop}%
\bibitem [{\citenamefont {Han}\ \emph {et~al.}(2003)\citenamefont {Han},
  \citenamefont {Logan}, \citenamefont {McElrath},\ and\ \citenamefont
  {Wang}}]{Han:2003wu}%
  \BibitemOpen
  \bibfield  {author} {\bibinfo {author} {\bibfnamefont {T.}~\bibnamefont
  {Han}}, \bibinfo {author} {\bibfnamefont {H.~E.}\ \bibnamefont {Logan}},
  \bibinfo {author} {\bibfnamefont {B.}~\bibnamefont {McElrath}}, \ and\
  \bibinfo {author} {\bibfnamefont {L.-T.}\ \bibnamefont {Wang}},\ }\href
  {\doibase 10.1103/PhysRevD.67.095004} {\bibfield  {journal} {\bibinfo
  {journal} {Phys. Rev.}\ }\textbf {\bibinfo {volume} {D67}},\ \bibinfo {pages}
  {095004} (\bibinfo {year} {2003})},\ \Eprint
  {http://arxiv.org/abs/hep-ph/0301040} {arXiv:hep-ph/0301040 [hep-ph]}
  \BibitemShut {NoStop}%
\bibitem [{\citenamefont {Kaplan}\ and\ \citenamefont
  {Schmaltz}(2003)}]{Kaplan:2003uc}%
  \BibitemOpen
  \bibfield  {author} {\bibinfo {author} {\bibfnamefont {D.~E.}\ \bibnamefont
  {Kaplan}}\ and\ \bibinfo {author} {\bibfnamefont {M.}~\bibnamefont
  {Schmaltz}},\ }\href {\doibase 10.1088/1126-6708/2003/10/039} {\bibfield
  {journal} {\bibinfo  {journal} {JHEP}\ }\textbf {\bibinfo {volume} {10}},\
  \bibinfo {pages} {039} (\bibinfo {year} {2003})},\ \Eprint
  {http://arxiv.org/abs/hep-ph/0302049} {arXiv:hep-ph/0302049 [hep-ph]}
  \BibitemShut {NoStop}%
\bibitem [{\citenamefont {Cortes~Maldonado}\ \emph {et~al.}(2012)\citenamefont
  {Cortes~Maldonado}, \citenamefont {Fernandez~Tellez},\ and\ \citenamefont
  {Tavares-Velasco}}]{CortesMaldonado:2011pi}%
  \BibitemOpen
  \bibfield  {author} {\bibinfo {author} {\bibfnamefont {I.}~\bibnamefont
  {Cortes~Maldonado}}, \bibinfo {author} {\bibfnamefont {A.}~\bibnamefont
  {Fernandez~Tellez}}, \ and\ \bibinfo {author} {\bibfnamefont
  {G.}~\bibnamefont {Tavares-Velasco}},\ }\href {\doibase
  10.1088/0954-3899/39/1/015003} {\bibfield  {journal} {\bibinfo  {journal} {J.
  Phys.}\ }\textbf {\bibinfo {volume} {39}},\ \bibinfo {pages} {015003}
  (\bibinfo {year} {2012})},\ \Eprint {http://arxiv.org/abs/1109.4390}
  {arXiv:1109.4390 [hep-ph]} \BibitemShut {NoStop}%
\bibitem [{\citenamefont {Oda}\ \emph {et~al.}(2015)\citenamefont {Oda},
  \citenamefont {Okada},\ and\ \citenamefont {Takahashi}}]{Oda:2015gna}%
  \BibitemOpen
  \bibfield  {author} {\bibinfo {author} {\bibfnamefont {S.}~\bibnamefont
  {Oda}}, \bibinfo {author} {\bibfnamefont {N.}~\bibnamefont {Okada}}, \ and\
  \bibinfo {author} {\bibfnamefont {D.-s.}\ \bibnamefont {Takahashi}},\ }\href
  {\doibase 10.1103/PhysRevD.92.015026} {\bibfield  {journal} {\bibinfo
  {journal} {Phys. Rev.}\ }\textbf {\bibinfo {volume} {D92}},\ \bibinfo {pages}
  {015026} (\bibinfo {year} {2015})},\ \Eprint
  {http://arxiv.org/abs/1504.06291} {arXiv:1504.06291 [hep-ph]} \BibitemShut
  {NoStop}%
\bibitem [{\citenamefont {Das}\ \emph {et~al.}(2016)\citenamefont {Das},
  \citenamefont {Oda}, \citenamefont {Okada},\ and\ \citenamefont
  {Takahashi}}]{Das:2016zue}%
  \BibitemOpen
  \bibfield  {author} {\bibinfo {author} {\bibfnamefont {A.}~\bibnamefont
  {Das}}, \bibinfo {author} {\bibfnamefont {S.}~\bibnamefont {Oda}}, \bibinfo
  {author} {\bibfnamefont {N.}~\bibnamefont {Okada}}, \ and\ \bibinfo {author}
  {\bibfnamefont {D.-s.}\ \bibnamefont {Takahashi}},\ }\href {\doibase
  10.1103/PhysRevD.93.115038} {\bibfield  {journal} {\bibinfo  {journal} {Phys.
  Rev.}\ }\textbf {\bibinfo {volume} {D93}},\ \bibinfo {pages} {115038}
  (\bibinfo {year} {2016})},\ \Eprint {http://arxiv.org/abs/1605.01157}
  {arXiv:1605.01157 [hep-ph]} \BibitemShut {NoStop}%
\bibitem [{\citenamefont {Olive}\ \emph {et~al.}(2014)\citenamefont {Olive}
  \emph {et~al.}}]{Agashe:2014kda}%
  \BibitemOpen
  \bibfield  {author} {\bibinfo {author} {\bibfnamefont {K.~A.}\ \bibnamefont
  {Olive}} \emph {et~al.} (\bibinfo {collaboration} {Particle Data Group}),\
  }\href {\doibase 10.1088/1674-1137/38/9/090001} {\bibfield  {journal}
  {\bibinfo  {journal} {Chin. Phys.}\ }\textbf {\bibinfo {volume} {C38}},\
  \bibinfo {pages} {090001} (\bibinfo {year} {2014})}\BibitemShut {NoStop}%
\bibitem [{\citenamefont {Osland}\ \emph {et~al.}(2010)\citenamefont {Osland},
  \citenamefont {Pankov},\ and\ \citenamefont {Tsytrinov}}]{Osland:2009dp}%
  \BibitemOpen
  \bibfield  {author} {\bibinfo {author} {\bibfnamefont {P.}~\bibnamefont
  {Osland}}, \bibinfo {author} {\bibfnamefont {A.~A.}\ \bibnamefont {Pankov}},
  \ and\ \bibinfo {author} {\bibfnamefont {A.~V.}\ \bibnamefont {Tsytrinov}},\
  }\href {\doibase 10.1140/epjc/s10052-010-1272-z} {\bibfield  {journal}
  {\bibinfo  {journal} {Eur. Phys. J.}\ }\textbf {\bibinfo {volume} {C67}},\
  \bibinfo {pages} {191} (\bibinfo {year} {2010})},\ \Eprint
  {http://arxiv.org/abs/0912.2806} {arXiv:0912.2806 [hep-ph]} \BibitemShut
  {NoStop}%
\bibitem [{\citenamefont {Hahn}(2001)}]{Hahn:2000kx}%
  \BibitemOpen
  \bibfield  {author} {\bibinfo {author} {\bibfnamefont {T.}~\bibnamefont
  {Hahn}},\ }\href {\doibase 10.1016/S0010-4655(01)00290-9} {\bibfield
  {journal} {\bibinfo  {journal} {Comput. Phys. Commun.}\ }\textbf {\bibinfo
  {volume} {140}},\ \bibinfo {pages} {418} (\bibinfo {year} {2001})},\ \Eprint
  {http://arxiv.org/abs/hep-ph/0012260} {arXiv:hep-ph/0012260 [hep-ph]}
  \BibitemShut {NoStop}%
\bibitem [{\citenamefont {Hahn}\ and\ \citenamefont
  {Perez-Victoria}(1999)}]{Hahn:1998yk}%
  \BibitemOpen
  \bibfield  {author} {\bibinfo {author} {\bibfnamefont {T.}~\bibnamefont
  {Hahn}}\ and\ \bibinfo {author} {\bibfnamefont {M.}~\bibnamefont
  {Perez-Victoria}},\ }\href {\doibase 10.1016/S0010-4655(98)00173-8}
  {\bibfield  {journal} {\bibinfo  {journal} {Comput. Phys. Commun.}\ }\textbf
  {\bibinfo {volume} {118}},\ \bibinfo {pages} {153} (\bibinfo {year}
  {1999})},\ \Eprint {http://arxiv.org/abs/hep-ph/9807565}
  {arXiv:hep-ph/9807565 [hep-ph]} \BibitemShut {NoStop}%
\bibitem [{\citenamefont {Bardin}\ and\ \citenamefont
  {Passarino}(1999)}]{Bardin:1999ak}%
  \BibitemOpen
  \bibfield  {author} {\bibinfo {author} {\bibfnamefont {D.~{\relax Yu}.}\
  \bibnamefont {Bardin}}\ and\ \bibinfo {author} {\bibfnamefont
  {G.}~\bibnamefont {Passarino}},\ }\href@noop {} {\emph {\bibinfo {title}
  {{The standard model in the making: Precision study of the electroweak
  interactions}}}}\ (\bibinfo {year} {1999})\BibitemShut {NoStop}%
\bibitem [{\citenamefont {Arbuzov}\ \emph {et~al.}(2006)\citenamefont
  {Arbuzov}, \citenamefont {Awramik}, \citenamefont {Czakon}, \citenamefont
  {Freitas}, \citenamefont {Grunewald}, \citenamefont {Monig}, \citenamefont
  {Riemann},\ and\ \citenamefont {Riemann}}]{Arbuzov:2005ma}%
  \BibitemOpen
  \bibfield  {author} {\bibinfo {author} {\bibfnamefont {A.~B.}\ \bibnamefont
  {Arbuzov}}, \bibinfo {author} {\bibfnamefont {M.}~\bibnamefont {Awramik}},
  \bibinfo {author} {\bibfnamefont {M.}~\bibnamefont {Czakon}}, \bibinfo
  {author} {\bibfnamefont {A.}~\bibnamefont {Freitas}}, \bibinfo {author}
  {\bibfnamefont {M.~W.}\ \bibnamefont {Grunewald}}, \bibinfo {author}
  {\bibfnamefont {K.}~\bibnamefont {Monig}}, \bibinfo {author} {\bibfnamefont
  {S.}~\bibnamefont {Riemann}}, \ and\ \bibinfo {author} {\bibfnamefont
  {T.}~\bibnamefont {Riemann}},\ }\href {\doibase 10.1016/j.cpc.2005.12.009}
  {\bibfield  {journal} {\bibinfo  {journal} {Comput. Phys. Commun.}\ }\textbf
  {\bibinfo {volume} {174}},\ \bibinfo {pages} {728} (\bibinfo {year}
  {2006})},\ \Eprint {http://arxiv.org/abs/hep-ph/0507146}
  {arXiv:hep-ph/0507146 [hep-ph]} \BibitemShut {NoStop}%
\bibitem [{\citenamefont {Erler}\ and\ \citenamefont
  {Ramsey-Musolf}(2005)}]{Erler:2004in}%
  \BibitemOpen
  \bibfield  {author} {\bibinfo {author} {\bibfnamefont {J.}~\bibnamefont
  {Erler}}\ and\ \bibinfo {author} {\bibfnamefont {M.~J.}\ \bibnamefont
  {Ramsey-Musolf}},\ }\href {\doibase 10.1103/PhysRevD.72.073003} {\bibfield
  {journal} {\bibinfo  {journal} {Phys. Rev.}\ }\textbf {\bibinfo {volume}
  {D72}},\ \bibinfo {pages} {073003} (\bibinfo {year} {2005})},\ \Eprint
  {http://arxiv.org/abs/hep-ph/0409169} {arXiv:hep-ph/0409169 [hep-ph]}
  \BibitemShut {NoStop}%
\bibitem [{\citenamefont {Han}\ \emph {et~al.}(2013)\citenamefont {Han},
  \citenamefont {Langacker}, \citenamefont {Liu},\ and\ \citenamefont
  {Wang}}]{Han:2013mra}%
  \BibitemOpen
  \bibfield  {author} {\bibinfo {author} {\bibfnamefont {T.}~\bibnamefont
  {Han}}, \bibinfo {author} {\bibfnamefont {P.}~\bibnamefont {Langacker}},
  \bibinfo {author} {\bibfnamefont {Z.}~\bibnamefont {Liu}}, \ and\ \bibinfo
  {author} {\bibfnamefont {L.-T.}\ \bibnamefont {Wang}},\ }\href@noop {} {\
  (\bibinfo {year} {2013})},\ \Eprint {http://arxiv.org/abs/1308.2738}
  {arXiv:1308.2738 [hep-ph]} \BibitemShut {NoStop}%
\bibitem [{\citenamefont {Langacker}\ and\ \citenamefont
  {Luo}(1992)}]{Langacker:1991pg}%
  \BibitemOpen
  \bibfield  {author} {\bibinfo {author} {\bibfnamefont {P.}~\bibnamefont
  {Langacker}}\ and\ \bibinfo {author} {\bibfnamefont {M.-x.}\ \bibnamefont
  {Luo}},\ }\href {\doibase 10.1103/PhysRevD.45.278} {\bibfield  {journal}
  {\bibinfo  {journal} {Phys. Rev.}\ }\textbf {\bibinfo {volume} {D45}},\
  \bibinfo {pages} {278} (\bibinfo {year} {1992})}\BibitemShut {NoStop}%
\bibitem [{\citenamefont {Gulov}\ and\ \citenamefont
  {Skalozub}(2000)}]{Gulov:2000eh}%
  \BibitemOpen
  \bibfield  {author} {\bibinfo {author} {\bibfnamefont {A.~V.}\ \bibnamefont
  {Gulov}}\ and\ \bibinfo {author} {\bibfnamefont {V.~V.}\ \bibnamefont
  {Skalozub}},\ }\href {\doibase 10.1007/s100520000496} {\bibfield  {journal}
  {\bibinfo  {journal} {Eur. Phys. J.}\ }\textbf {\bibinfo {volume} {C17}},\
  \bibinfo {pages} {685} (\bibinfo {year} {2000})},\ \Eprint
  {http://arxiv.org/abs/hep-ph/0004038} {arXiv:hep-ph/0004038 [hep-ph]}
  \BibitemShut {NoStop}%
\bibitem [{\citenamefont {Pevzner}(2018)}]{Pevzner:2018chl}%
  \BibitemOpen
  \bibfield  {author} {\bibinfo {author} {\bibfnamefont {A.~O.}\ \bibnamefont
  {Pevzner}},\ }\href@noop {} {\bibfield  {journal} {\bibinfo  {journal}
  {Nonlin. Phenom. Complex Syst.}\ }\textbf {\bibinfo {volume} {21}},\ \bibinfo
  {pages} {30} (\bibinfo {year} {2018})},\ \Eprint
  {http://arxiv.org/abs/1803.07508} {arXiv:1803.07508 [hep-ph]} \BibitemShut
  {NoStop}%
\end{thebibliography}%
\end{document}